\documentclass[11pt, traditabstract]{aa}  
\graphicspath{ {./images/} }
\usepackage{graphicx}
\usepackage[varg]{txfonts}
\usepackage{longtable}
\usepackage{ulem}
\usepackage{threeparttable} 
\titlerunning{V CVn Photopolarimetry}
\authorrunning{Neilson et al} 

\begin{document} 


   \title{A Multi-Year Photopolarimetric Study of the Semi-Regular Variable V~CVn and Identification of Analogue Sources}

   \author{H. Neilson\inst{\ref{inst1}}\
          \and
          N. Steenken\inst{\ref{inst2}}
          \and
          J. Simpson\inst{\ref{inst3}}
          \and
          R. Ignace\inst{\ref{inst4}}
          \and
          M. Shrestha\inst{\ref{inst5}, \ref{inst6}}
          \and
          C. Erba\inst{\ref{inst4}}
          \and
          G. Henson\inst{\ref{inst4}}
          }

   \institute{Department of Physics \& Physical Oceanography, Memorial University of Newfoundland \& Labrador, St.~John's, NL, Canada A1C~5S7\label{inst1}\\
              \email{hneilson@mun.ca}
            \and
            Sternwarte Freimann, 80939 Munich, Germany\label{inst2}
        \and
            Parton, Scotland, UK\label{inst3}
            \and
            Department of Physics \& Astronomy, East Tennessee State University, Johnson City, TN 37614, USA\label{inst4} 
            \and
            Astrophysics Research Institute, Liverpool John Moores University, Liverpool Science Park, 146 Brownlow Hill, Liverpool, L3 5RF, UK\label{inst5} 
             \and
             Steward Observatory, University of Arizona, 933 North Cherry Avenue, Tucson, AZ 85721-0065, USA \label{inst6} \\
             }
             
   \date{}
\abstract{
The semi-regular variable star V~Canum Venaticorum (V CVn) is well-known for its unusual linear polarization position angle (PA). Decades of observing V CVn reveal a nearly constant PA spanning hundreds of pulsation cycles. This phenomenon has persisted through variability that has ranged by 2 magnitudes in optical brightness and through variability in the polarization amplitude over 0.3\% and 6.9\%. Additionally, the polarization fraction of V~CVn varies inversely with brightness.

This paper presents polarization measurements obtained over three pulsation cycles. We find that the polarization maximum does not always occur precisely at the same time as the brightness minimum. Instead, we observe a small lead or lag in relation to the brightness minimum, spanning a period of a few days up to three weeks. Furthermore, the PA sometimes exhibits a non-negligible rotation, especially at lower polarization levels.  

To elucidate the unusual optical behavior of V~CVn, we present a list of literature sources that also exhibit polarization variability with a roughly fixed PA. We find this correlation occurs in stars with high tangential space velocities, i.e., ``runaway'' stars, suggesting that the long-term constant PA is related to how the circumstellar gas is shaped by the star's high-speed motion through the interstellar medium. 
}



    \maketitle
%

\section{Introduction}

Semi-regular variable stars (SRVs) are intriguing objects. They are evolved, pulsating red giant stars that reflect the future of our own Sun. SRVs are distinguished from Mira-type stars by their smaller brightness amplitude that is less than 2.5 mag(V).  In particular, SRV stars of type 'a' (SRa) show an identifiable and sustainable primary pulsation period and in many cases secondary pulsation periods. 

SRV stars, which are bright and so have been observed for centuries, are recognized as important standard candles that also follow a number of period-luminosity relations \citep{Trabucchi2017, Lebzelter2019}. This correspondence contributes to our understanding of their evolution, and makes SRVs valuable to extragalactic and cosmological studies. Even after so many observations and studies SRVs continue to present a number of questions, including the role of convection, stellar and circumstellar magnetic fields, atmospheric dynamics, winds, along with how these stars interact with their circumstellar medium \citep{1992A&A...263...97K, 2019Ap.....62..556Kn}.

One of the more surprising questions arises from linear polarization measurements of some SRV stars. \citet{Kruszewski1968} and \citet{Shawl1975b, Shawl1975a} presented early polarization measurements for V~CVn (HD 115898). This M4e-M6eIIIa late AGB star has a highly variable polarization and is located at a distance of 501~pc \citep{GaiaCollaboration2020}. Linear polarization measurements of unresolved sources can measure deviations from circular symmetry as projected onto the sky. Thus, the detection of linear polarization from a star may indicate an asymmetry that due to the presence of convection, wind bow shocks, or disks \citep{Shrestha2021}.

\cite{Kruszewski1968} presented time-domain polarization measurements of V~CVn showing that the polarization fraction changes over the approximate 194~day primary pulsation period, while the position angle remains constant. What is more curious is that the polarization fraction appears to be anti-correlated to the brightness of the star: when the star is brightest, the polarization is at a minimum; and when the star is dimmest, the polarization is at a maximum. This surprising behaviour was confirmed to be a function of wavelength by \cite{Magalhaes1986b,Magalhaes1986a}. While still unexplained, this phenomenon is not unique to V CVn.
 
\cite{Kruszewski1968} identified similar polarization variability in the star L$_2$~Puppis. This result added to the challenge of understanding these two stars, but resolved polarimetric observations provided information about the circumstellar structure of the L$_2$~Pup system. \cite{Kervella2015} used VLT SPHERE/ZIMPol polarimetric observations to discover the presence of a circumstellar disk. This disk is remarkable as its asymmetry projected on the sky explains the constant position angle of the unresolved linear polarization, but \emph{not} the variation of the polarization itself.  
 
\cite{Neilson2014} reviewed time-domain polarization measurements for V~CVn, and confirmed the periodic variability of the polarization while the position angle stays approximately constant.  They also confirmed that the maximum value of the polarization appeared to correspond to minimum brightness, and the minimum polarization to maximum brightness.  They suggested that the polarization and position angle time-domain observations could be explained by a pulsation-driven, spherically symmetric dusty wind that interacts with an asymmetric stellar wind bow shock.  The presence of a constant bow shock could explain the constant position angle.  When the star approaches maximum brightness, it also approaches maximum radius and minimum effective temperature. Thus, dust will form in the spherically symmetric outflow, and that dust will be more likely to interact with scattering photons from the star. This will cause the polarization to decrease to a minimum. Conversely, as the shell expands, the density decreases and the shell will interact with the bow shock, creating a highly asymmetric system, and resulting in a maximum polarization value, as shown by \cite{Shrestha2021}.

An alternative model was proposed by \cite{Safonov2019} based on differential speckle polarimetric observations of V~CVn. They  introduced the idea of scattering arcs or reflection nebulae in the circumstellar envelope of the star; hereafter we refer to these as simply "blobs".  In their model, some blobs are in the background, and some are in the foreground of the system.  In this geometry, the authors argued that the polarization variability could be explained if the star undergoes a form of non-radial variability of its brightness which could be either dipolar pulsation or rotational variation, such that when the star appears brightest, an observer behind the star would see V~CVn at maximum brightness, and vice versa.  Because one blob appears to be behind the star, when we see V~CVn at minimum light, the distant blob will see intercept photons from the star at maximum brightness on the far side, and thus there will be more backscattered photons, hence the polarization will be greatest.  
 
The polarization variability of V~CVn can thus be explained by both of these scenarios, but they each depend on the circumstellar medium (CSM) structures of V~CVn and L$_2$~Pup. Understanding the CSM of these stars offers important insights into how evolved semi-regular variable stars lose mass, and potentially connects to their future evolution into planetary nebulae. 
In a collaboration between professional and amateur astronomers, we consider here new time-domain polarization observations of V~CVn (led by co-authors Steenken and Simpson) that offer new tests of the two models mentioned.
This is an especially valuable collaboration because V~CVn is too bright for most large telescopes. Therefore it was advantageous to perform this investigation in cooperation with amateur astronomers.  In Section~\ref{obs}, we discuss the tools and methods used to obtain new observations of V~CVn. In Section~\ref{results}, we compare the polarimetric observations with the visual brightness of the star. Section~\ref{discussion} then presents a review of potential analogues to V~CVn and L$_2$~Pup, and we suggest that these stars are actually the prototypes of a class of polarimetric semi-regular variable stars (``V~CVn type" stars). We discuss the various  CSM and polarizations models for V~CVn to discern if and how these new observations might provide new insights into this system. Finally, Section~\ref{concl} presents our conclusions and  suggestions for future observations.

\section{Observations and Data Reduction} \label{obs}

Optical polarization observations and photometry of V~CVn were carried out independently by two of the co-authors (Steenken and Simpson; hereafter ``observers'') operating Schmidt-Cassegrain-Telescopes (SCT). The observers are located in Munich/Germany (Steenken, Ob1) and in Parton/Scotland (Simpson, Ob2). Both telescopes have Dual-Beam-Polarimeters with V-band filters and cooled CCD cameras mounted at the Cassegrain focus. Customary SCTs were used, with OB1 having a focal length of 2800 mm and aperture of 280 mm, and OB2 using a two different SCTs with focal lengths of 1500 and 2000 mm and apertures of 150 and 200 mm (Ob2) on a German equatorial mount. Object centering and autoguiding was achieved with separate guide scopes and guiding cameras.

\begin{figure*}[t]
\centering
\includegraphics[scale=0.64]{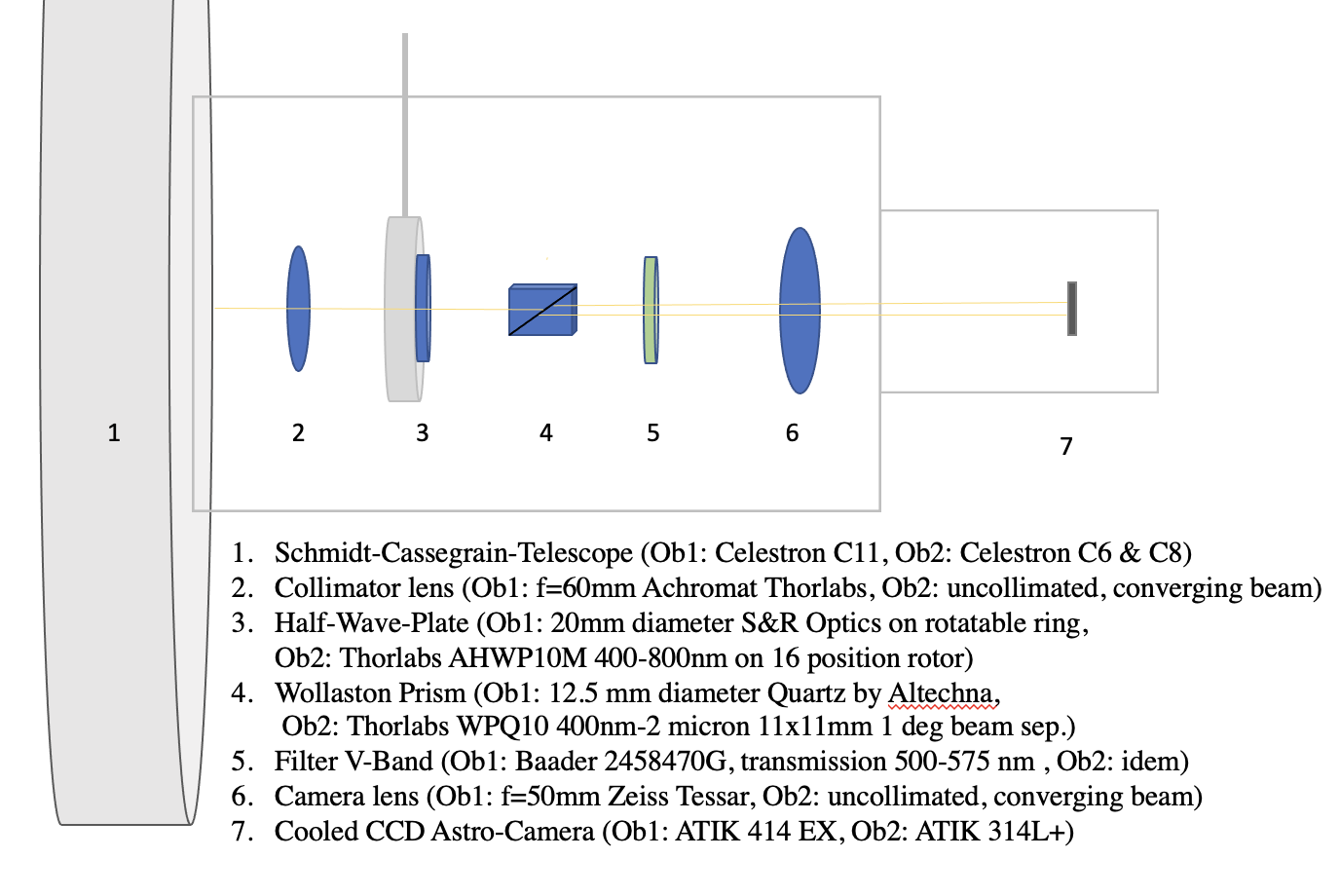}
\caption{\label{fig:polarimeter}
Optical design of the dual-beam-polarimeter, with components along the optical path identified and labeled.}
\end{figure*}

Both polarimeters were built and calibrated in 2018-2019 to measure the linear optical polarization ($P$) of stars brighter than 9 mag, with an error of $\pm 0.1$ \%  polarization and $\pm 5$ degrees of polarization position angle (PA) in the relevant polarization range of V~CVn in the past i.e. 0.3\% - 6.9\%.
An achromatic Half-Wave-Plate (HWP) was manually rotated into the positions 0, 22.5, 45 and 67.5 degrees to rotate the polarization plane 0, 45, 90 and 135 degrees. The 0~degree position is aligned with the celestial North direction. The instrumental 0 deg position has been nominally aligned with the celestial NS reference and offsets from this are checked during each observing session, using measurements obtained either by plate solving a stacked measurement image and/or by disabling tracking so as to generate an EW trace on measured images and therefore a NS reference. These offsets are removed during data processing. Calibration, later performed, uses high polarisation standards to correct instrumental position angle offsets arising from any half-wave plate/Wollaston prism misalignments. A Wollaston prism splits of the star into a parallel and a perpendicular beam. The beams are passed through a V-band filter and focused onto a CCD detector of $1392 \times 1040$ pixels for a cooled 16-bit Astro camera. The optical design is shown in Fig.~\ref{fig:polarimeter}.

\begin{figure}
\centering
\includegraphics[width=\columnwidth]{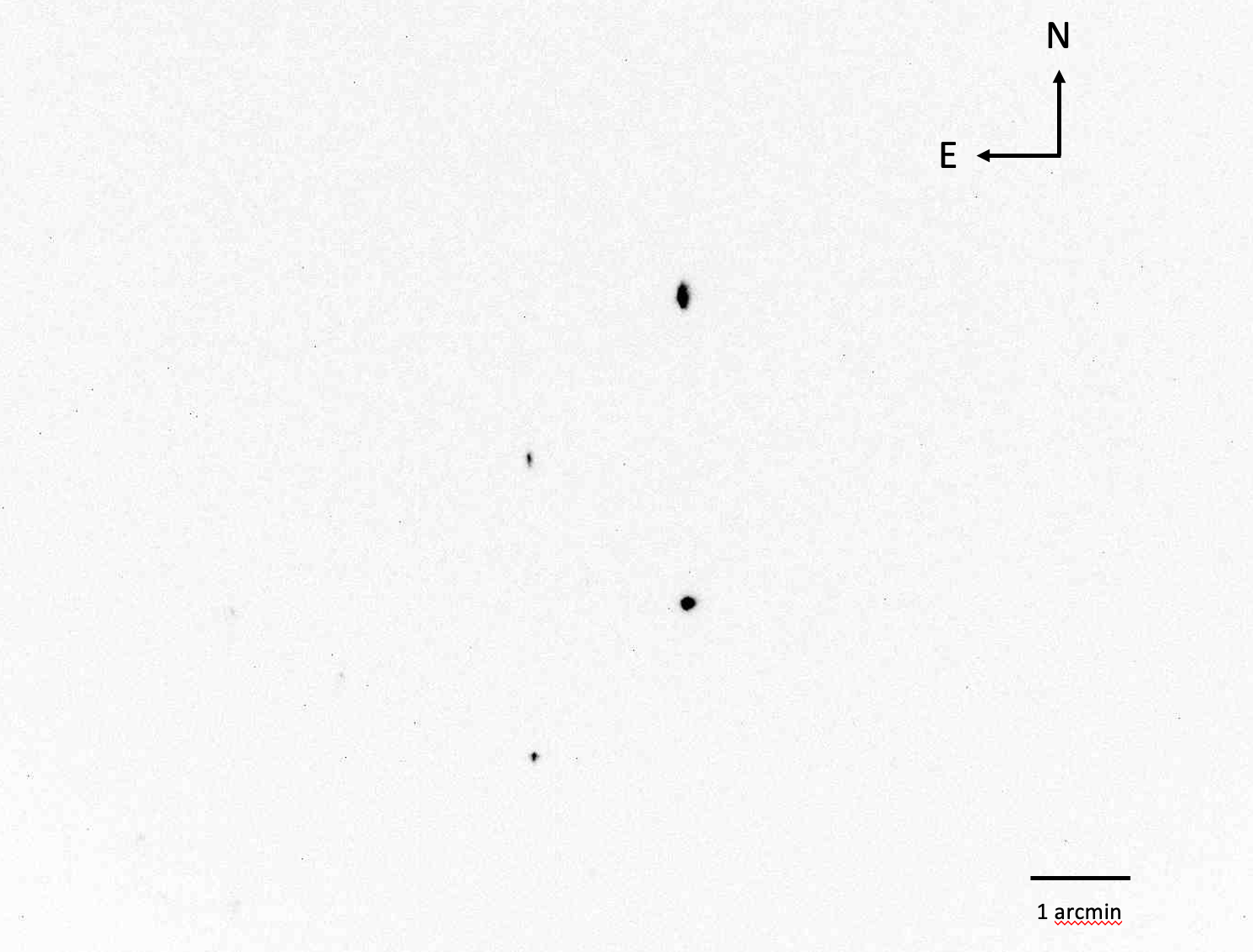}
\caption{\label{fig:vcvn_img}
An example of one image of V CVn split into parallel (bright star at top) and perpendicular beams (bright star at bottom) from 2022-07-03 at 0 degree rotation position. The star left below V CVn is BD+46 1863. }
\end{figure}

The images were captured and analyzed with standard image capturing and photometry software (for Ob1: AstroArt~6\footnote[1]{www.msb-astroart.com} and for Ob2: AstroImageJ\footnote[2]{www.astro.louisville.edu}).
The exposure times vary from 2 to 15 seconds depending on the star brightness, weather conditions and telescope. For one observation, 40 to 100 images in each rotation position of the HWP were exposed. Batch processing was used for the dark field correction, stellar alignment, and photometry of the parallel and perpendicular stellar images. Aperture radii selection, based on the image's full width at half maximum [FWHM] -- typically 1.5-2 times FWHM for the signal and 3 resp. 5 times FWHM respectively for inner/outer annuli for sky background evaluation -- provided a flux measurement that is robust against seeing variations, giving confidence in the method even when the object images were slightly distorted. Signal-to-noise ratios (SNR) of over 400 of each image taken as calculated by the photometry software were achieved, subject to atmospheric conditions. Fig.~\ref{fig:vcvn_img} shows an example of an image made by Ob 1. 

Fluxes of parallel $f_{par_i}$ and perpendicular beams $f_{perp_i}$ were measured for every image $i$ with the photometry software. Polarization ($P$), including correction for bias, and Position Angle (PA) were calculated as described in \cite{Clarke2010}. For the error analysis, the study by \cite{Patat_2006} was used.

To calculate the measured Stokes $Q_{m}$ and $U_{m}$ parameters, parallel and perpendicular flux measurements are combined in the following way, for angle positions $\theta$:

\begin{eqnarray}
    N_\theta & = & \frac{\sum{f_{par_i}} - \sum{f_{perp_i}}}{\sum{f_{par_i}} + \sum{f_{perp_i}}} , \\ 
Q_{m} & = & \frac{1}{2}\{N_{0} -N_{45}) ~ {\rm and}\\
U_{m} & = & \frac{1}{2}\{N_{22.5} -N_{67.5})
\end{eqnarray}

\noindent where $N_\theta$ is the fractional photon
count between parallel and perpendicular beams at
the specified angel of the Half-Wave-Plate.

Any optical design of a polarimeter has to deal with instrumental polarization caused by the telescope or the optical components in the polarimeter, including the filter. To remove the instrumental polarization effects, the Stokes parameters of unpolarized stars in Table~\ref{tab1} were measured as zero-point offsets and calibrated Stokes parameters calculated using:
\begin{table}
\caption{Unpolarized stars used as standards, and their polarization according to \cite{Schmidt1992}\label{tab1}}
\centering
\begin{tabular}{l  c  c  c  c  c  c} 
 \hline
 Star & $mag(V)$ &$P$ (\%) & $\sigma_P$ (\%) & Observer \\ \hline \hline
 $\beta$~Cas & 2.3 & 0.037 & 0.024& 1,2\\
 $\theta$~UMa & 3.2 & 0.037 & 0.014& 1,2\\ 
 $\beta$~UMa & 2.4& 0.009 & 0.019 &2\\
 HD 21447 & 5.1 & 0.051 & 0.020& 1\\ 
 \hline
\end{tabular}
\end{table}
\begin{eqnarray}
Q_{cal} & = & Q_{m} -Q_{0} ,~{\rm and} \\
U_{cal} &= & U_{m} -U_{0}
\end{eqnarray}

Calibration standard high polarization stars like $\phi$~Cas and 2H~Cam were measured in agreement to the published values of \cite{Hsu1982} as shown in Table~\ref{tab2}.
\begin{table}[!ht]
\caption{Highly polarized stars measurements and comparison to reference} \label{tab2}
    \centering
    \begin{tabular}{ l l l l l }
    \hline
        Star & JD & $P_{cal}$  & $PA_{cal}$ & Ob  \\ \hline\hline
        $\phi$~Cas & 2459077 & 3.30\% & 98.8 & 1 \\ \hline
        $\phi$~Cas & 2459280 & 3.42\% & 95.7 & 1 \\ \hline
        $\phi$~Cas & 2459448 & 3.32\% & 95.6 & 1 \\ \hline
        $\phi$~Cas & 2459452 & 3.12\% & 96.6 & 1 \\ \hline
        $\phi$~Cas & 2459599 & 3.35\% & 93.8 & 1 \\ \hline
        $\phi$~Cas & 2459599 & 3.34\% & 92.7 & 1 \\ \hline
        $\phi$~Cas & 2459622 & 3.29\% & 92.8 & 1 \\ \hline
        $\phi$~Cas & 2459808 & 3.44\% & 96.0 & 1 \\ \hline
        $\phi$~Cas & Average  & 3.32\% & 95.3 & 1 \\ \hline
        ~ & ~ & ~ & ~ & ~ \\ \hline
        $\phi$~Cas & 2459287 & 3.41\% & 93.3 & 2 \\ \hline
        $\phi$~Cas & 2459287 & 3.24\% & 92.0 & 2 \\ \hline
        $\phi$~Cas & 2459314 & 3.44\% & 92.7 & 2 \\ \hline
        $\phi$~Cas & Average  & 3.36\% & 92.7 & 2 \\ \hline
        $\phi$~Cas & \cite{Hsu1982} & 3.34\% & 92.3 & ~ \\ \hline
        ~ & ~ & ~ & ~ & ~ \\ \hline
        2H Cam & 2459319 & 3.37\% & 115.2 & 2 \\ \hline
        2H Cam & 2459543 & 3.53\% & 115.0 & 2 \\ \hline
        2H Cam & Average   & 3.45\% & 115.1 & 2 \\ \hline
        2H Cam & \cite{Hsu1982} & 3.49\% & 116.6 \\ \hline
    \end{tabular}
\end{table}
Both the instrumental set-ups and zero-point offsets were kept unchanged during the observing seasons from January until September to ensure comparable results. Calibration stars were observed each season to determine the stability and the reliability of the polarization position angle.

V-band photometry of V~CVn was obtained directly before or after the polarimetric measurements with nearby comparison stars. Ob1 selected nearby star HD~116957, with $mag(V) = 5.87$, as comparison. Ob2 used an ensemble of the four AASVO comparison stars \footnote[3]{Chart X28069ADU, AAVSO Variable Star Database}.

The measured polarization is composed of the intrinsic polarization of V CVn and the interstellar polarization.

The star BD+46 1863 is located only about 2 arcmin south-west of V CVn at a distance of about 1670 pc and was used estimate the interstellar polarization. The measurements performed on the nearby stars TYC 3460-1345-1 and TYC 3460-2045-1 gave values of polarization between 0 and 0.5\%. However, due to the high measurement uncertainties, these measurements were not used. The estimate is as shown in table 3 and is the basis for the calculation of the intrinsic polarization of V CVn.

\begin{table}[!ht]
\caption{Estimate of interstellar polarization based on nearby star BD+46 1863 \label{tab4}}
    \centering
    \begin{tabular}{l c c c c c}
    \hline
        $P_{is}$ & $PA_{is}$ & $Q_{is}$ & $U_{is}$ & $\sigma_Q$ & $\sigma_U$\\ \hline\hline
        0.34\% & 123 & -0.14\% & -0.31\% & 0.010\% & 0.009\% \\ \hline
    \end{tabular}
\end{table}

Intrinsic Stokes $Q_{intr}$ and $U_{intr}$ parameters were calculated according to:

\begin{eqnarray}
Q_{intr} & = & Q_{cal} -Q_{is} ,~{\rm and}\\
U_{intr} &= & U_{cal} -U_{is}
\end{eqnarray}

\noindent Intrinsic Polarization $P_{intr}$ and intrinsic polarization angle $PA_{intr}$ according to:

\begin{eqnarray}
P_{intr} & = & \sqrt{Q_{intr}^2+U_{intr}^2},
~{\rm and} \\
PA_{intr} & = & \frac{1}{2}\,\tan^{-1} \left(\frac{U_{intr}}{Q_{intr}}\right).
\end{eqnarray}

The standard error of the mean values of Stokes Q, U were calculated according to \cite{Clarke2010} using the student-t distribution for 95\% confidence limits. Errors of  P and PA and were calculated as follows:

\begin{equation}
\sigma_P=\frac{\sqrt{Q^2\sigma_{Q}^2+U^2\sigma_U^2}~{\rm }}{P}
\end{equation}

\begin{equation}
\sigma_{PA}=\frac{\sqrt{Q^2\sigma_U^2+U^2\sigma_Q^2}}{2P^2} 
\end{equation}

In the data reduction process polarization has been de-biased using the \cite{Wardle1974} formulation:
\begin{equation}
P_{debias}=\sqrt{P_{cal}^2 - \sigma_P^2 }
\end{equation}

\begin{figure}
\centering
\includegraphics[width=\columnwidth]{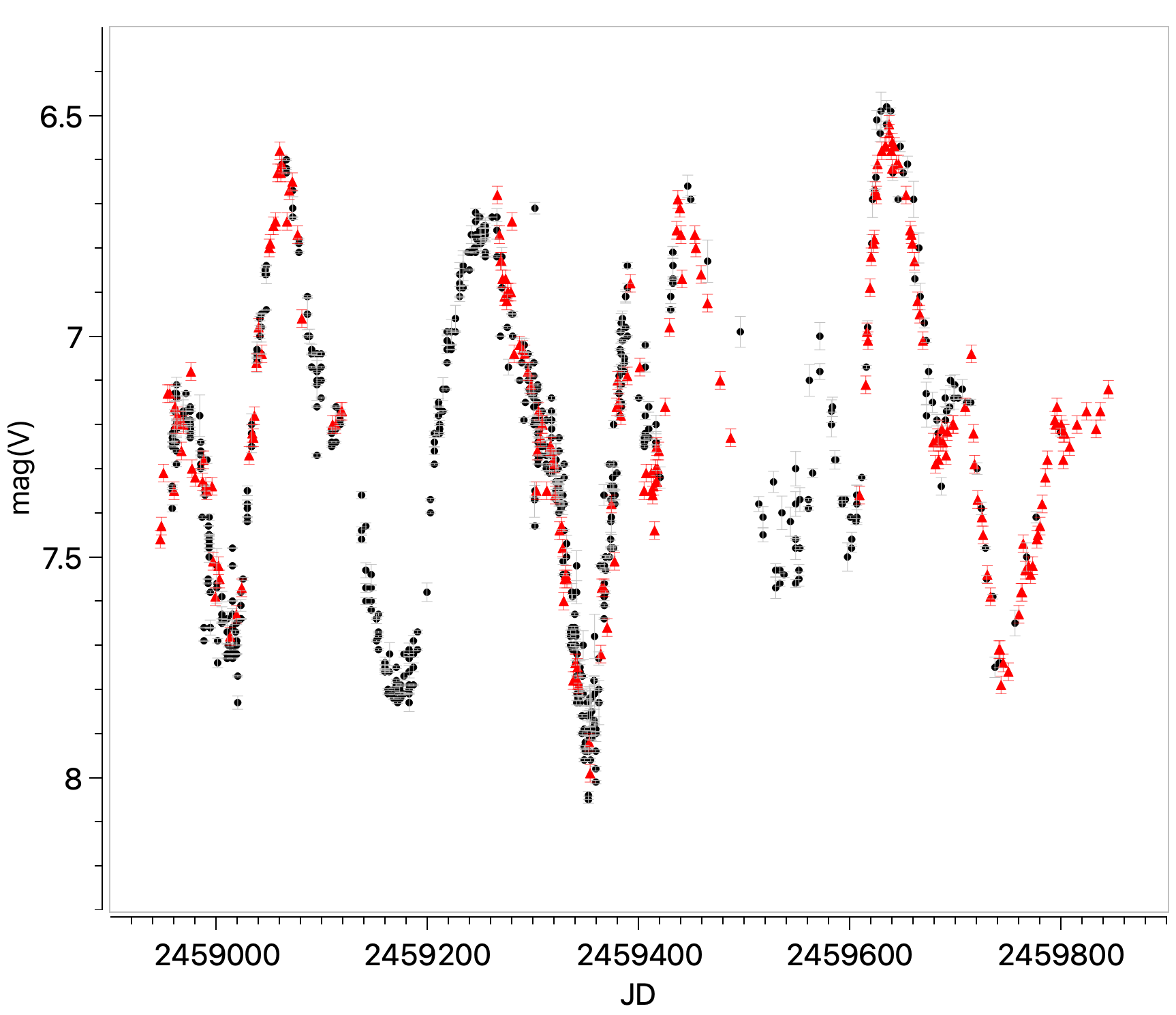}
\caption{\label{fig:puls_cycl}
The V-band light curve of V~CVn from 2020-2022. The red triangles are the photometry measurements provided by Ob1 and Ob2. The grey dots are for AASVO data (CCD measurements only). The three pulsation cycles covered by polarization measurements in this study in red, namely at the left pulsation cycle 1/2020 in the middle 1/2021 and at the right cycle 1/2022. It can be seen that the data from AASVO agrees well with the data measured by Ob1 and Ob2. Pulsation cycle 1/2021 showed the lowest brightness of all minima considered.}
\end{figure}

A total of 227 measurements were taken on 51 nights in 2020, 89 nights in 2021, and 87 nights in 2022, with a table of measurements in Appendix~\ref{AppendB}\footnote[4]{All raw data can be obtained from https://zenodo.org/record/7997101}. During this span of time, V~CVn went through 5 pulsation cycles, as shown in Fig.~\ref{fig:puls_cycl}. The polarization measurements covered the first pulsation cycle in 2020, 2021 and 2022 (hereafter 1/2020, 1/2021 and 1/2022).

\section{Results} \label{results}

\begin{figure*}
    
\centering

\includegraphics[scale=0.47]{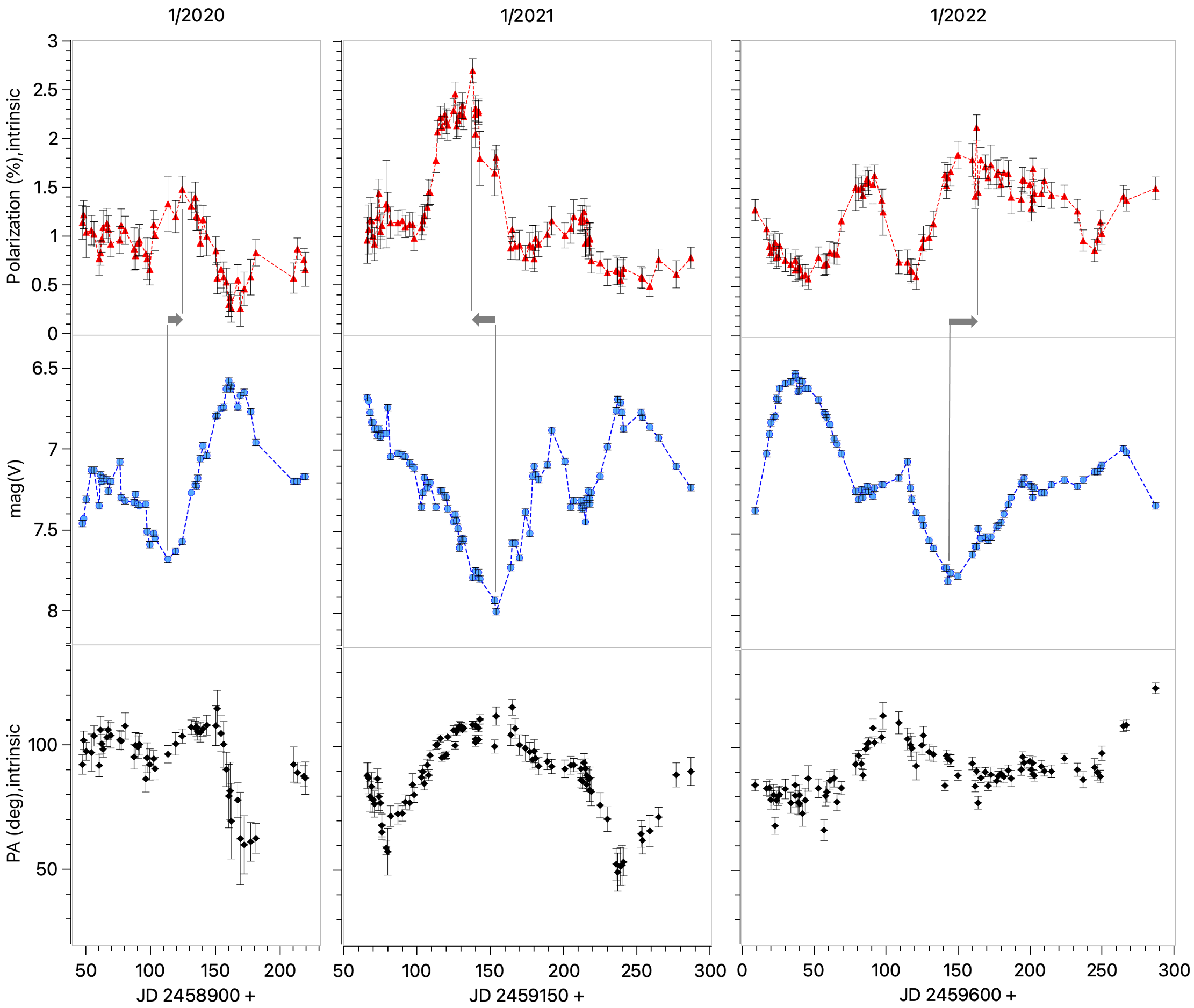}

\caption{\label{fig:Cycles20-22}
\textbf{Photopolarimetry of the observed pulsation cycles}. Intrinsic polarization curves in red are shown in the upper row. The light curves in blue are shown in the middle row. The intrinsic position angle (PA) is shown in the lower row as black.
 In all three pulsation cycles polarization and photometry are generally anti-correlated, especially true around times of brightness maxima and minima. However, the curves are not exactly anti-correlated as indicated by the grey arrows.
A phase shift between polarization and total light is most clearly visible around the brightness minimum of pulsation cycle 1/2021, but also seen in pulsation cycles 1/2020 and 1/2022. Comparison between the 3 cycles reveals that the polarization curve can either lead or lag the light curve.
The intrinsic PA ranges generally within 80 -- 120 degrees, except when the star is brighter than 7 mag(V) and/or the polarization drops significantly below 0.7\%. At such times, the PA can drop below 50 degrees. }
\end{figure*}

The relationship between intrinsic polarization and brightness over the course of all three pulsation cycles is shown in Fig.~\ref{fig:Cycles20-22}. 
The polarization in the V-Band is highly variable, and is generally anti-correlated with brightness. The larger polarizations were measured around brightness minima, and are consistent with earlier studies by \cite{Serkowski2001} and \cite{Davidson2014}.

Thanks to the high cadence of measurements covering three pulsation cycles, the temporal relationship between the star's light curve and polarization can be analyzed in more detail. Pulsation cycle 1/2021 showed the strongest obscuration associated with the highest measured intrinsic polarization of all three observed pulsation cycles.
Of particular interest is the observation that the maximum polarization does not occur exactly at the time of the minimum brightness of V CVn. We observe that in the three cycles that the polarization maximum can lead the brightness minimum by several weeks as well as lag it. Especially in cycle 1/2021 polarization led by about 16 days and in cycle 1/2022 polarization lagged again by about 20 days. This lead/lag phenomenon was observed around brightness minima and is indicated by a grey arrows in Fig.~\ref{fig:Cycles20-22}.  The pulsation cycle 1/2020 is shown in the left column of the figures. Polarisation maximum of 1.47\% was reached at JD 2459024. The brightness at 7.68 mag(V) minimum occurred 11 days (JD 2459013) before. Polarization was lagging the light curve.
In the pulsation cycle 1/2021 (middle column) polarization was leading the light curve. The maximum of intrinsic polarization  of 2.70\% was measured at JD 2459338 16 days before the first minimum of the light curve which occured at JD 2459354 and reached 7.99 mag(V). In this pulsation cycle, the star darkened significantly more and reached much higher polarization values than in the other two cycles. In the pulsation cycle 1/2022 (right column) polarization was lagging again the light curve. Second light minimum occurred on JD 2459743 at 7.79 mag(V) which was followed by the polarization maximum of 2.11\% on JD 2459763 20 days later. Due to the irregular course of the light curve and the daily variations in brightness and polarization, an exact determination of the shift is not possible. Polarization showed strong variations while going through the maximum. There are indications that phase shifts between the curves may occur when the polarization changes significantly within a few weeks. During periods of little change in brightness, there appears to be little correlation between the curves.  The position angle PA stays generally within a range of 80 - 120 degrees, except when the intrinsic polarization drops below 0.7\%.

%


\section{Discussion} \label{discussion}

\subsection{A New Class of V~CVn-type Stars} \label{new_class}
The {\it Combined General Catalog of Variable Stars} \citep{Samus2004} lists 348 SRa and 1201 SRb stars, yet polarimetric measurements have only been published for a few of them. As a consequence, it is unknown whether significant variation of the polarization during a pulsation cycle of more than 1\% is typical or exceptional for semi-regular variables. 

\begin{figure}
\centering
\includegraphics[scale=0.42]{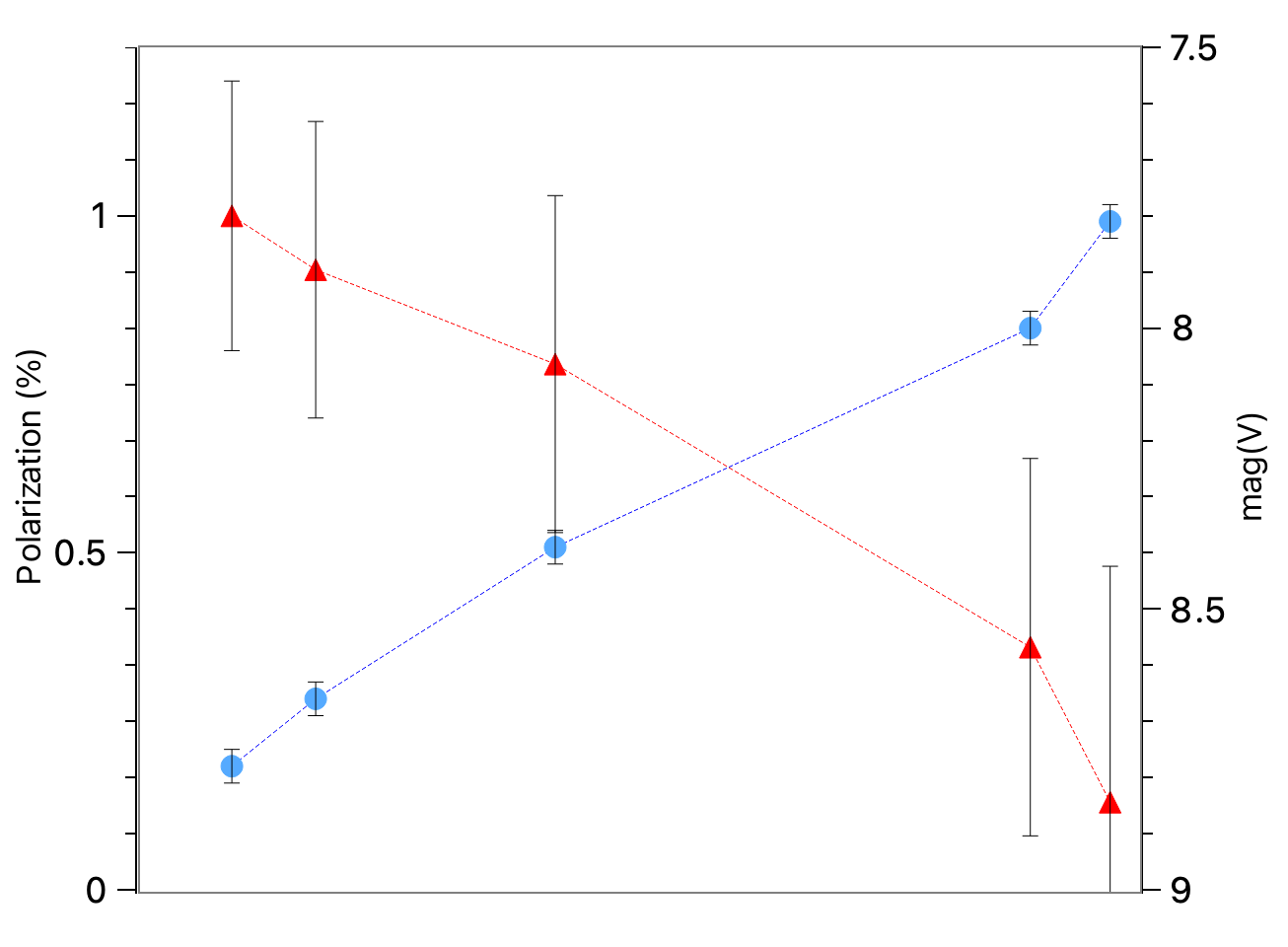}
 \caption{\label{fig:ZUma}
Polarization measurements (red triangles) and photometry (blue  dots) of Z~UMa in 2021 (Ob1). Five epochs of photopolarimetric measurements were obtained. The data give indications of  an anti-correlation between the measured polarization and brightness.}
\end{figure}

While V~CVn is not the only semi-regular evolved star with significant variable polarization, it by far has the greatest number of polarization measurements obtained since the 1960s. Table~\ref{tab:var_stars} (\ref{AppendA}) lists the basic stellar parameters of five other semi-regular stars that show polarimetric behavior similar to V~CVn itself. The following list summarizes their observed properties:
\begin{itemize}
    \item UZ Ari (IRC +20052) appears similar to V~CVn in many respects. It is situated at high galactic latitude ($-$31~deg) at a similar distance (539~pc), and has a tangential velocity of 59~km/sec. The period of UZ~Ari is 163 days, about 30 days shorter than of V~CVn. UZ~Ari has a spectral class of M8, with an apparent brightness that is significantly lower than V~CVn, and varies between 11.8 and 12.6 mag(V). \citet{Baug2014} were able to measure the diameter of the star in K-band during three lunar occultations. They found values of 4.5 - 6.0 $\pm$ 0.5~mas without a correlation to phase. The few published polarization measurements in the V-band vary between 2 and 3.5\%, with a decade long stable PA around 130 deg \citep{Kruszewski1976,Baug2014}. If future polarization measurements of UZ~Ari confirm the stable range of PA, then the physical origin of the asymmetry could be the same as in V~CVn. 
    
    \item AK~Peg has a very similar period of variability and spectral type to V~CVn. At its distance of 1329 pc, the star ranges in brightness between 8.6 - 10.2 mag(V). The few polarization measurements published by \citet{Serkowski2001} indicated polarization values between 1.5 and 3.4\%, a fairly large intrinsic level of polarimetric variability similar to UZ~Ari.
    
    \item RX~Boo has variabilty that is classified as an SRb, with a distance of 156~pc. Using the Infrared Optical Array Imaging Interferometer (IOTA), \citet{Ragland2006} measured a stellar diameter of 17.5~mas in H-Band that showed no asymmetry.  Unfortunately, only a few older polarization measurements are available for this star \citep{Vartanyan1968}, with values ranging from 0.4 to 1.9\%. The change in polarization is comparable
    to UZ~Ari and AK~Peg.
    
    \item Z~UMa is an SRb star at a distance of 296 pc. \citet{Dyck1971} published polarization measurements ranging between 0.2 and 1.8\%.  Measurements at 5 epochs by Ob1 in 2021 (Fig.~\ref{fig:ZUma}) confirmed variable polarization from 0.13 to 1.0\%, showing an anti-correlation with brightness.  Again, the amplitude of variable polarization is comparable to the previous three objects.
    
    \item L$_2$~Pup is an SRb star on the southern hemisphere. The polarization measurements of \citet{Magalhaes1986a} reveal a high level of polarization variabilty between 0 and 8\%, accompanied by a steady PA of 165-180~degrees. If the few historical polarization V-band measurements are compared with photometric data from the AASVO, the polarization appears anti-correlated with brightness level. This level of variable polarization is considerably larger than the four preceding stars, but is comparable to V~CVn.
\end{itemize}
 
Due to a distance of only 56 pc, and the superior imaging performance of VLT SPHERE/ZIMPol, it has been possible to detect a dust disk responsible for the asymmetry of L$_2$~Pup \citep{Kervella2015}. For the other stars listed above, the reasons for the asymmetries remain unclear. None of these stars are located in the galactic plane, with its higher interstellar density, but some of them show high tangential or radial velocities.
 
\begin{figure}
\centering
\includegraphics[scale=0.30]{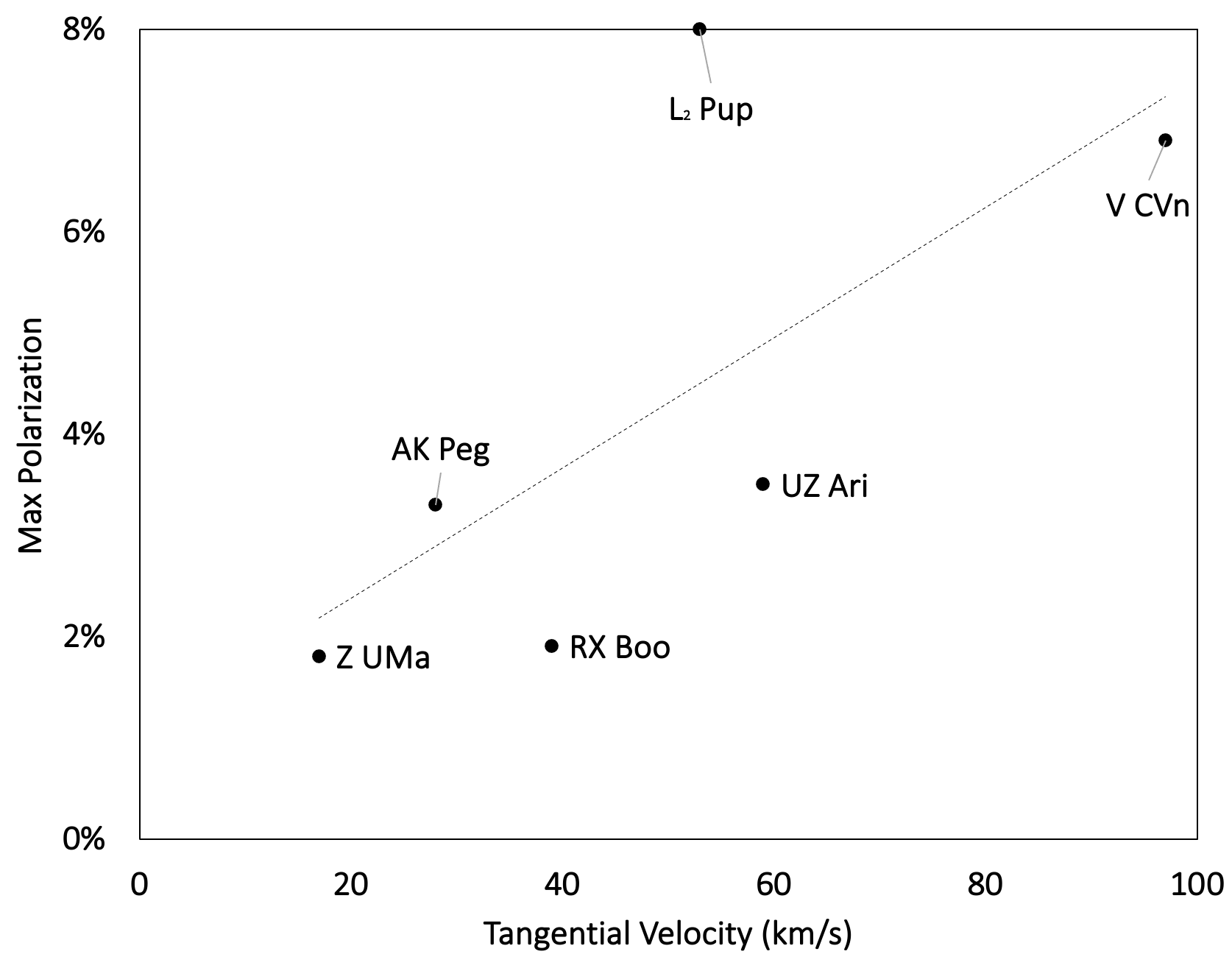}
\caption{ \label{fig:tangVel_Corr}
Maximum polarization vs. tangential velocity of the stars in Table~\ref{tab:var_stars}, indicating an higher maximum polarization with increased tangential velocity.}
\end{figure}
 
In Fig.~\ref{fig:tangVel_Corr}, we see hints of a relationship between the tangential velocity and the observed maximum polarization. Stars with significant tangential velocities and slow winds, such as those from cool, evolved stars, would be expected to host a bow shock along the plane of the sky. That bow shock, if unresolved, would create a significant asymmetry on the plane of the sky with an associated polarization signature \citep{Shrestha2021}. The amount of polarization would depend on both the shape and density of the bow shock, along with the properties of the stellar wind.  
\citet{Shrestha2021} showed that polarization is a function of the type of dust in the bow shock, the density of the local interstellar medium, and the wind properties of the stars in question, so the apparent correlation of Figure~\ref{fig:tangVel_Corr} could be consistent with the model proposed by \cite{Neilson2014}. The polarization would be variable if, during the phase where V~CVn is brightest, the star ejects a dusty shell whose density is sufficient to wash out the bow shock signal and make the system appear more symmetric \citep{Shrestha2021}.

%
%
%
 
\cite{Safonov2019} argued that V~CVn cannot have a bow shock, due to a low ISM density expected from its high galactic latitude. However, the star is located only a few arcmin away from a molecular cloud discovered by \citet{Reach1994}. The molecular cloud G107.4~+70.9 is one of the densest clouds in that region, and lies roughly at the same distance as V~CVn, according to a study from \citet{Gladders1998}. It is plausible that the interstellar density near the fast moving V~CVn could be large enough to allow the formation of a bow shock.


\subsection{Possible models of the polarimetric variability} \label{models}

The polarimetric variability of V~CVn is surprising and counterintuitive. When the star reaches maximum light, the polarization approaches a minimum, and vice-versa. One might expect that if the source of the polarization were constant -- which would yield a constant position angle -- then at maximum light, the number of polarized photons would also be at a maximum. However, since the polarization is a normalized quantity, one would find that the polarization is also constant.
Satisfying the observations requires either of the following scenarios:
\begin{enumerate}
    \item The source of the polarization interacts with a different number of photons relative to what the observer sees \citep{Safonov2019}; or
    \item The source of the polarization changes, yet maintains the same deviation of symmetry on the sky \citep{Neilson2014}.
\end{enumerate}
 
\noindent The results of this work impose an additional requirement that a viable model should explain observed time differences between the polarization maximum and the light minimum.
 
If we consider the first option, \citet{Safonov2019} suggest that one of the dusty ``blobs'' is behind the star relative to the observer (c.f., Sec.~1). They suggest that if the light variation is not due to radial pulsation, and instead is some type of dipolar variability or rotational variability, then when the observer measures minimum light, the other side of the star appears at maximum light -- hence the blob ``sees" the greatest number of polarizing photons that get backscattered to the observer.  This model is intriguing, but requires the stellar variability to be asymmetric and non-radial. The pulsation amplitude of V~CVn is about 2 mag(V), and there are no known stars that pulsate with this amplitude non-radially. If this were a rotational phenomena, combining a period of about $\approx 194$~d with an assumed stellar radius of order 100~$R_\odot$ implies the rotation rate is about 25~km~s$^{-1}$, and is the same order as the critical rotation rate for the star.  

An additional challenge for this model is the apparent lead and lag time between polarization and flux. If the issue was only that the polarization maximum lagged behind the luminosity, one might explain it as a light-time delay of $t_{d} = 2 2d_{b}/c$, where $d_b$ is the distance between the star and blob. \citet{Safonov2019} reported a radius of the blobs around V~CVn of 35~$\pm$~1~mas. Using the latest distance data from \citet{GaiaCollaboration2020} of 501~pc, this results in a distance of the bipolar cloud of 17.5~AU. The time delay caused by the light travelling between the assumed bipolar cloud and the star is then 2.5~hr, and thus by far not enough to explain the observed time difference in our observational data between the light and polarization curves of up to two weeks. 

\citet{Neilson2014} suggest that the variable polarization is caused by a pulsation-driven dusty wind shell \citep[see also][]{Willson2000, Hofner2022} that collides with a stellar wind bow shock, which is expected since V~CVn is a runaway star.  They argue that the shell is driven outwards from the star near maximum light, and because the shell is symmetric, the polarization is then at a minimum. However, the position angle will remain constant because of the constant presence of the asymmetric bow shock around the shell.  As the shell expands and its density decreases, the polarization is set more by the near-constant density bow shock. This result is consistent with the apparent connection between the polarization amplitude and the tangential velocity observed for V~CVn and its potential analogues, as described in Section~\ref{new_class}. However, this model does not explain the apparent lead/lag time.

Additionally, instead of an asymmetry due to the presence of bow shocks, the wind may simply be asymmetric and variable.  Simulations performed by \citet{Aronson2017} using non-spherical circumstellar shells that contain clumped material predicted significant net polarization signals. If the maximum polarization occurs near the minimum light then the wind, at that phase, could be most asymmetric, and at maximum light and minimum polarization, the wind would be most symmetric. It is worth noting that this argument is independent of wind density -- since the polarization observed for V~CVn is normalized -- and is thus only concerned with changes in the symmetry of the wind as projected against the sky.  

Asymmetric stellar winds are rare, but can be caused by rapid rotation as suggested for Be supergiants \citep[e.g.,][]{Granada2010, Georgy2011}, magnetic fields as seen in hot, massive stars with strong magnetic fields \citep[e.g.][]{Erba2021,Subramanian2022}, or perhaps convection as suggested for massive evolved red supergiant stars \citep[e.g.,][]{Kaminsiki2019,Lopez2022}. There is no evidence that V~CVn is a rapidly rotating star. However, given its evolved stage of evolution, its critical rotational velocity is small. Rotation might be important, and could be consistent with a small population of variable-polarization evolved stars. Furthermore, one would expect the rotation to be slow; however, it is possible there were dynamic events between the progenitors of these stars, and a companion spun up the progenitor \citep[e.g.,][]{Staritsin2022},  while a supernova explosion or another dynamic event ejected the companion \citep{Dorigo2020}. In that case, the progenitor would evolve into a runaway red giant star with rapid rotation. While this scenario may be possible, it is not obvious how to test this idea, and the proposed scenario should be a rare occurrence at best.

Strong magnetic fields ($> 1kG$) are not common among evolved stars \citep{Grunhut2010}; however, strong magnetic fields in hot stars are known to shape their stellar winds and make them asymmetric \citep[e.g.][]{2002ApJ...576..413U}.  In hot (O- and B-type) stars, the magnetic field tends to be nearly dipolar, while in evolved stars the magnetic field is more tangled, and similar to that of the Sun. It is not clear that magnetic fields in this case can impact the wind structure in a meaningful manner, but the existence of a maser \citep{Wolak2012} around V~CVn implies that follow-up measurements would be a valuable test, and could be related to the observed polarization lead/lag. 

Convection is unlikely to be the main source of polarizing photons in these red giant stars.  Convection is not constant, and will create spots of some lifetime over different parts of a star that is not periodic. As such, the variable polarization should not be correlated with the brightness variation. 

\begin{figure}
\centering
\includegraphics[width=\columnwidth]{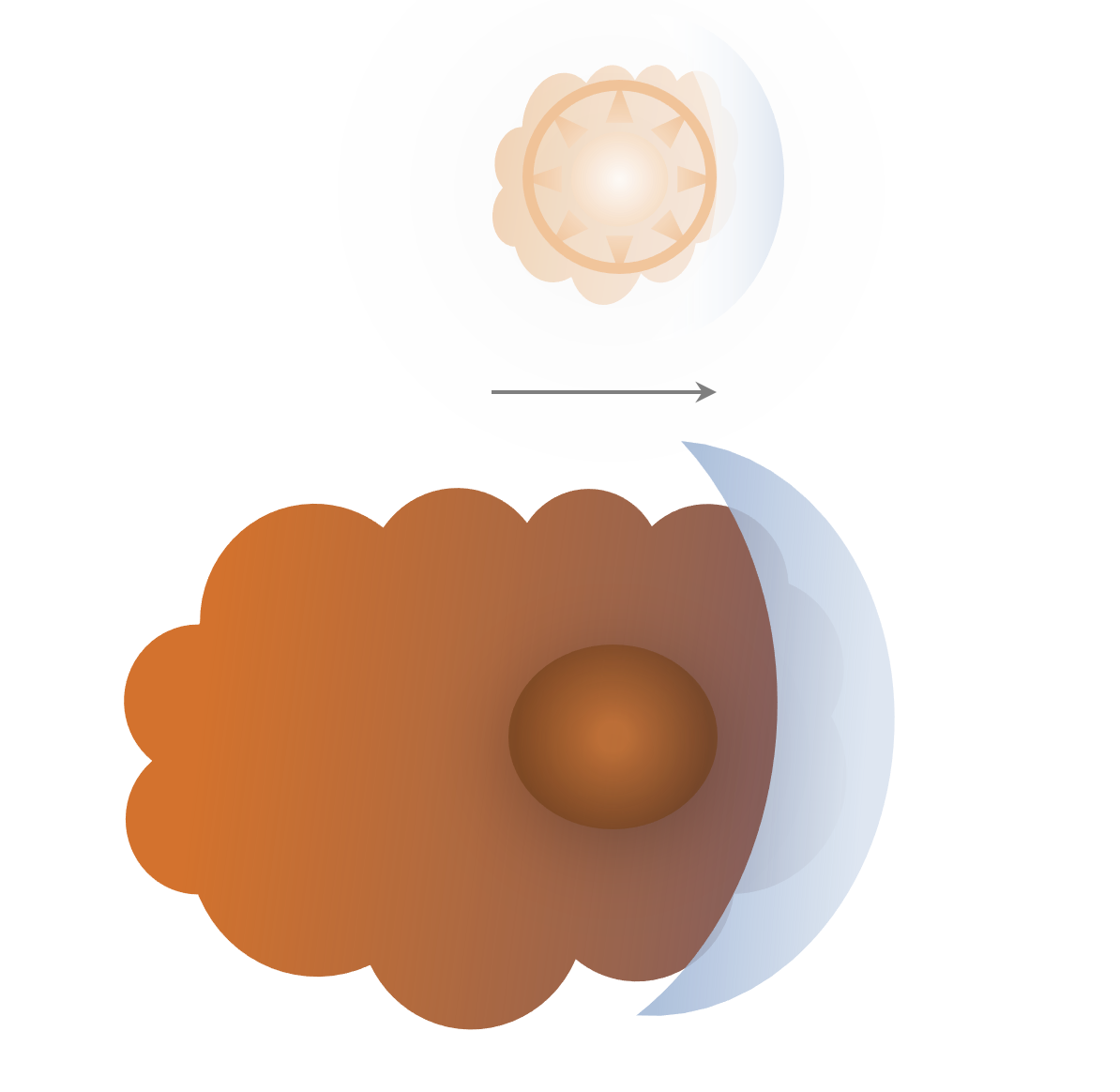}
\caption{ \label{fig:BowShock}
A schematic representation of the interaction between the pulsating star and the bow shock. {\bf Top:} at maximum brightness and minimum polarization, the dust is accelerated from the photosphere in a dense shell (circle) and the mass-loss rate is greatest. The star moves relative to the interstellar medium in the direction of the arrow. The wind shell is denser than the bow shock (light blue), and dominates the fraction of scattered light. {\bf Bottom:} Around minimum light there is less radiative acceleration, and dust forms in the photosphere, while the mass-loss rate is much smaller. The shell formed at the previous
flux maximum has expanded asymmetrically. The cloud-like structure of the dust envelope is supposed to show turbulence. A bow shock (blue) dominates the observed polarization, and is incorporated into the smaller density CSM. A lead/lag of the polarization and the light-curve can occur if convection causes the dust density in the line of sight and in the bow shock to develop differently. }
\end{figure}
 
Although convection cannot be the main source of polarization, it is possible that the polarization lead/lag is caused by convection. It is known that, for pulsating Cepheids, convection can perturb their regular light curves to vary cycle-to-cycle \citep{Derekas2012, Neilson2014a, Derekas2017}, and this is a potential cause of pulsation instability in variable red giant and supergiant stars. For V~CVn, it is likely that convection and granulation make the times of maximum and minimum light unstable and variable in the same way.  However, if the stellar wind is due to a dust shell being formed in the photosphere, then the wind is correlated more with the variability of the stellar radius and NOT the variability of the stellar luminosity. In Fig.~\ref{fig:BowShock}, we have sketched the relationship between the pulsating star and the bow shock. We assume that with each pulsation cycle a different pattern of convection cells is created. In a pulsation cycle -- where the dust is most dense first along the line of sight and only delayed in the direction of the bow shock -- the brightness minimum occurs before the polarization maximum, and the polarization curve lags behind the brightness curve, as observed in pulsation cycle 1/2020, and vice-versa as in cycle 1/2021.

\section{Conclusions}
\label{concl}

The new high-cadence photopolarimetry of V~CVn presented here raises new questions about the nature of the polarization variability both of this star and other semi-regular variable stars. In particular, we find that the polarization maximum does not occur at the same time as light minimum, and that time will be a lead or a lag. This phenomenon is not naturally explained by the models proposed by either \cite{Neilson2014} or \cite{Safonov2019}. Our investigation supports the hypothesis that the variable polarization is caused by a bow shock, and that the observed phase-shifts between polarization and light curves are caused by convective processes in the stellar photosphere. As such, we find that more high-cadence polarization measurements and simulations of V~CVn-type stars would be important if we wish to understand the relationship between their stellar properties, their motions, and their circumstellar media.


\begin{acknowledgements}
The authors gratefully thank the Referee for the constructive comments and recommendations which helped to improve the paper.
We acknowledge with thanks the variable star observations from the AAVSO International Database contributed by observers worldwide and used in this research. We thank Zenodo.org for hosting our data (https://zenodo.org/record/7997101).

This work has made use of data from the European Space Agency (ESA) mission
{\it Gaia} (\url{https://www.cosmos.esa.int/gaia}), processed by the {\it Gaia}
Data Processing and Analysis Consortium (DPAC,
\url{https://www.cosmos.esa.int/web/gaia/dpac/consortium}). Funding for the DPAC
has been provided by national institutions, in particular the institutions
participating in the {\it Gaia} Multilateral Agreement. MS is supported by an STFC consolidated grant number (ST/R000484/1) to LJMU. HN acknowledges funding from NSERC.

\end{acknowledgements}

\bibliographystyle{aa} 
\bibliography{vcvn_bib_2} 

\begin{appendix}
\section{The sample variable polarization semi-regular variable stars.}\label{AppendA}
\begin{table*}[!ht]
\begin{threeparttable}[b]
\caption{\label{tab:var_stars}
V CVn type stars: semi-regular variables with significant variable polarization. }
\centering
    {\renewcommand{\arraystretch}{1.5}
    \begin{tabular}{|l|c|c|c|c|c|c|}
    \hline 
         & V CVn & UZ Ari & AK Peg & RX Boo & Z UMa & L$_2$~Pup \\ \hline \hline
        Spectral Type\tnote{1}  & M4e-M6eIIIa & M8 & M4-M8e & M6.5e-M8IIIe & M5IIIe & M5IIIe-M6IIIe \\ \hline
        Galactic latitude ($^\circ$) & 70 & -31 & -43 & 69 & 57 & -15 \\ \hline
        Period (d)\tnote{1} & 194 & 163 & 194 & 162 & 195 & - \\ \hline
        $\Delta mag(V)$\tnote{1}  & 2.0 & 0.8 & 1.6 & 2.7 & 3.2 & 3.6 \\ \hline
        Polarimetric data (yr) &  1966\tnote{2} - 2022\tnote{2} &  1972\tnote{3} /2013\tnote{4} & 1967\tnote{5} & 1967\tnote{5} , 1970\tnote{6} & 1966\tnote{7} , 1970\tnote{6} &  1967\tnote{8}  -1982\tnote{9} \\ \hline
        Polarization (\%)\tnote{10} &  0.3 - 6.9 & 2 - 3.5 &  1.3 - 3.3 &  0.4 - 1.9 &  0.2 - 1.8 &  0 - 8 \\ \hline
        Postion angle ($^\circ$)\tnote{11} &  80 - 120 & 126 - 132 &  21 - 50 &  40 - 153 &  8 - 132 & 165 - 180 \\ \hline
        Stability in $\Delta PA$\tnote{12}  & yes & yes & unknown & unknown & unknown & yes \\ \hline
        Distance  (pc)\tnote{13} & 501 & 539 & 1329 & 156 & 296 & 56 \\ \hline
        $\mu$ (mas yr$^{-1}$)\tnote{13} & 40 & 23 & 45 & 53 & 12 & 340 \\ \hline
        $V_{\rm t}$ (km s$^{-1}$)\tnote{13} & 95 & 59 & 28 & 39 & 17 & 90 \\ \hline
        $V_{\rm r}$ (km s$^{-1}$)\tnote{13} & -47 & n.a. & -4 & -9 & -53 & 53 \\ \hline
    \end{tabular}
}
 \begin{tablenotes}
       \item [1] General catalog of variable stars \cite{GCVS2017}
       \item [2] \cite{Kruszewski1968}, \cite{Dyck1971}, \cite{Serkowski2001}, \cite{Davidson2014}, This study
       \item [3] \cite{Kruszewski1976}
       \item [4] \cite{Baug2014}
       \item [5] R.A. Vartanyan, Intrinsic Polarization of Stellar Light From RX Boo, AB Cyg, Astrofyzika, 1968
       \item [6] \cite{Dyck1971}
       \item [7] \cite{Serkowski1966}
       \item [8] \cite{Kruszewski1968}
       \item [9] \cite{Magalhaes1986}
       \item [10] Range of measured polarization in V-Band as published in the above mentioned studies (RX Boo: intrinsic P)
       \item [11] Range of measured polarization angles in V-Band as published in the above mentioned studies
       \item [12] Yes, if position angle remains constant over the period of available polarization data
       \item [13] \cite{GaiaCollaboration2020}
     \end{tablenotes}
  \end{threeparttable}
\end{table*}

\section{The recorded photometric measurements of V~CVn}\label{AppendB}
\begin{table*}[!ht]
\caption{\label{tab:journal_2020}
Photopolarimetry journal of pulsation cycle 1/2020.}
\centering
    {\renewcommand{\arraystretch}{1.1}
    \begin{tabular}{|ccccccccccc|}
    \hline
        JD & $P_{cal}$ & $PA_{cal}$ & $Q_{cal}$ & $\sigma_Q$ & $U_{cal}$ & $\sigma_{U}$ & $mag(V)$ & $\sigma_{mag(V)}$ & $Exp.$ & $Ob.$ \\
        (d) & (debiased)& ($^\circ$) & & & & &  & &(sec)&\\
        \hline\hline
        2458947.3 & 0.0133 & 98.6 & -0.0128 & 0.0018 & -0.0039 & 0.0013 & 7.46 & 0.02 & 8 & 1 \\ \hline
        2458948.3 & 0.0149 & 106.1 & -0.0126 & 0.0014 & -0.0080 & 0.0013 & 7.43 & 0.02 & 8 & 1 \\ \hline
        2458950.3 & 0.0127 & 103.3 & -0.0114 & 0.0010 & -0.0057 & 0.0010 & 7.31 & 0.02 & 10 & 1 \\ \hline
        2458954.3 & 0.0127 & 102.8 & -0.0117 & 0.0024 & -0.0056 & 0.0027 & 7.13 & 0.02 & 10 & 1 \\ \hline
        2458956.3 & 0.0130 & 108.3 & -0.0105 & 0.0011 & -0.0078 & 0.0011 & 7.13 & 0.02 & 10 & 1 \\ \hline
        2458960.3 & 0.0097 & 100.7 & -0.0091 & 0.0007 & -0.0036 & 0.0008 & 7.35 & 0.02 & 10 & 1 \\ \hline
        2458961.3 & 0.0112 & 110.8 & -0.0085 & 0.0014 & -0.0075 & 0.0013 & 7.16 & 0.02 & 10 & 1 \\ \hline
        2458962.3 & 0.0124 & 106.0 & -0.0105 & 0.0007 & -0.0066 & 0.0010 & 7.2 & 0.02 & 10 & 1 \\ \hline
        2458963.3 & 0.0133 & 103.8 & -0.0119 & 0.0008 & -0.0062 & 0.0008 & 7.18 & 0.02 & 10  & 1 \\ \hline
        2458966.3 & 0.0140 & 107.5 & -0.0115 & 0.0015 & -0.0081 & 0.0010 & 7.19 & 0.02 & 10 & 1 \\ \hline
        2458967.3 & 0.0136 & 109.9 & -0.0105 & 0.0012 & -0.0088 & 0.0010 & 7.26 & 0.02 & 10 & 1 \\ \hline
        2458969.3 & 0.0120 & 108.8 & -0.0095 & 0.0007 & -0.0073 & 0.0015 & 7.2 & 0.02 & 10 & 1 \\ \hline
        2458976.3 & 0.0123 & 107.3 & -0.0102 & 0.0008 & -0.0070 & 0.0009 & 7.08 & 0.02 & 10 & 1 \\ \hline
        2458977.3 & 0.0137 & 106.3 & -0.0116 & 0.0008 & -0.0074 & 0.0012 & 7.3 & 0.02 & 10 & 1 \\ \hline
        2458980.3 & 0.0135 & 111.2 & -0.0101 & 0.0009 & -0.0092 & 0.0020 & 7.32 & 0.05 & 10 & 1 \\ \hline
        2458987.3 & 0.0110 & 102.5 & -0.0100 & 0.0010 & -0.0047 & 0.0012 & 7.33 & 0.02 & 10 & 1 \\ \hline
        2458988.3 & 0.0105 & 106.5 & -0.0089 & 0.002 & -0.0058 & 0.0009 & 7.28 & 0.02 & 10 & 1 \\ \hline
        2458990.3 & 0.0119 & 105.2 & -0.0103 & 0.0009 & -0.006 & 0.0012 & 7.34 & 0.02 & 10 & 1 \\ \hline
        2458991.3 & 0.0121 & 105.9 & -0.0104 & 0.0009 & -0.0064 & 0.0012 & 7.35 & 0.02 & 10 & 1 \\ \hline
        2458996.3 & 0.0097 & 96.0 & -0.0095 & 0.0012 & -0.002 & 0.0012 & 7.34 & 0.02 & 10 & 1 \\ \hline
        2458997.3 & 0.0100 & 103.0 & -0.009 & 0.0011 & -0.0044 & 0.0014 & 7.51 & 0.02 & 10 & 1 \\ \hline
        2458999.4 & 0.0085 & 102.1 & -0.008 & 0.0019 & -0.0036 & 0.0012 & 7.59 & 0.02 & 10 & 1 \\ \hline
        2459002.4 & 0.0133 & 100.5 & -0.0125 & 0.0012 & -0.0048 & 0.0009 & 7.52 & 0.02 & 10 & 1 \\ \hline
        2459003.4 & 0.0119 & 97.8 & -0.0115 & 0.0009 & -0.0032 & 0.0013 & 7.55 & 0.02 & 10 & 1 \\ \hline
        2459013.4 & 0.0155 & 101.2 & -0.0144 & 0.0011 & -0.0059 & 0.0014 & 7.68 & 0.02 & 10 & 1 \\ \hline
        2459019.4 & 0.0144 & 105.1 & -0.0126 & 0.0028 & -0.0073 & 0.0014 & 7.63 & 0.02 & 10 & 1 \\ \hline
        2459024.4 & 0.0175 & 107.0 & -0.0146 & 0.0013 & -0.0098 & 0.0012 & 7.57 & 0.02 & 10 & 1 \\ \hline
        2459031.4 & 0.0161 & 110.3 & -0.0122 & 0.0008 & -0.0105 & 0.001 & 7.27 & 0.02 & 10 & 1 \\ \hline
        2459034.4 & 0.0170 & 109.8 & -0.0131 & 0.001 & -0.0108 & 0.001 & 7.22 & 0.02 & 10 & 1 \\ \hline
        2459035.4 & 0.0151 & 110.6 & -0.0114 & 0.0011 & -0.01 & 0.0013 & 7.23 & 0.02 & 10 & 1 \\ \hline
        2459036.4 & 0.0148 & 109.1 & -0.0116 & 0.0011 & -0.0092 & 0.0011 & 7.18 & 0.02 & 10 & 1 \\ \hline
        2459038.4 & 0.0122 & 109.8 & -0.0094 & 0.0007 & -0.0078 & 0.001 & 7.06 & 0.02 & 10 & 1 \\ \hline
        2459040.4 & 0.0147 & 110.1 & -0.0112 & 0.001 & -0.0095 & 0.0009 & 6.98 & 0.02 & 10 & 1 \\ \hline
        2459043.4 & 0.0130 & 111.6 & -0.0095 & 0.0011 & -0.0089 & 0.001 & 7.04 & 0.02 & 10 & 1 \\ \hline
        2459050.4 & 0.0114 & 112.0 & -0.0083 & 0.0012 & -0.008 & 0.0025 & 6.8 & 0.05 & 10 & 1 \\ \hline
        2459051.4 & 0.0089 & 117.7 & -0.0051 & 0.001 & -0.0074 & 0.0011 & 6.79 & 0.02 & 10 & 1 \\ \hline
        2459054.4 & 0.0094 & 110.8 & -0.0071 & 0.0012 & -0.0063 & 0.0014 & 6.75 & 0.02 & 10 & 1 \\ \hline
        2459056.4 & 0.0083 & 108.5 & -0.0068 & 0.0016 & -0.0051 & 0.0016 & 6.74 & 0.02 & 10 & 1 \\ \hline
        2459058.4 & 0.0073 & 102.5 & -0.0067 & 0.0011 & -0.0031 & 0.0009 & 6.63 & 0.02 & 10 & 1 \\ \hline
        2459060.4 & 0.0046 & 102.7 & -0.0042 & 0.0009 & -0.002 & 0.001 & 6.58 & 0.02 & 10 & 1 \\ \hline
        2459061.4 & 0.0053 & 101.0 & -0.005 & 0.0006 & -0.002 & 0.0013 & 6.63 & 0.02 & 10 & 1 \\ \hline
        2459062.3 & 0.0034 & 101.1 & -0.0034 & 0.0011 & -0.0014 & 0.0009 & 6.61 & 0.02 & 10 & 1 \\ \hline
        2459067.3 & 0.0064 & 93.7 & -0.0064 & 0.001 & -0.0008 & 0.001 & 6.74 & 0.02 & 10 & 1 \\ \hline
        2459069.3 & 0.0026 & 98.8 & -0.0029 & 0.0015 & -0.0009 & 0.0012 & 6.67 & 0.02 & 10 & 1 \\ \hline
        2459072.3 & 0.0034 & 83.3 & -0.0036 & 0.0015 & 0.0009 & 0.0016 & 6.65 & 0.02 & 10 & 1 \\ \hline
        2459077.3 & 0.0047 & 78.9 & -0.0045 & 0.0011 & 0.0018 & 0.0016 & 6.77 & 0.02 & 10 & 1 \\ \hline
        2459081.3 & 0.0071 & 74.4 & -0.0062 & 0.0013 & 0.0037 & 0.0017 & 6.96 & 0.02 & 10 & 1 \\ \hline
        2459110.4 & 0.0079 & 76.8 & -0.0071 & 0.0008 & -0.0035 & 0.0011 & 7.2 & 0.02 & 20 & 2 \\ \hline
        2459113.4 & 0.0103 & 82.4 & -0.0101 & 0.0011 & -0.0027 & 0.0009 & 7.2 & 0.02 & 20 & 2 \\ \hline
        2459118.4 & 0.0093 & 82.5 & -0.009 & 0.0002 & -0.0024 & 0.0007 & 7.17 & 0.02 & 30 & 2 \\ \hline
        2459119.4 & 0.0082 & 81.9 & -0.0079 & 0.0003 & -0.0023 & 0.0012 & 7.17 & 0.02 & 30 & 2 \\ \hline
 \hline
    \end{tabular}
    }
\end{table*}

\begin{table*}[!ht]
\caption{\label{tab:journal_2021-1}
Journal of measurements from 2021, part 1.}
\centering
    {\renewcommand{\arraystretch}{1.1}
    \begin{tabular}{|ccccccccccc|}
    \hline
    JD & $P_{cal}$ & $PA_{cal}$ & $Q_{cal}$ & $\sigma_Q$ & $U_{cal}$ & $\sigma_{U}$ & $mag(V)$ & $\sigma_{mag(V)}$ & $Exp.$ & $Ob.$ \\
        (d) & (debiased)& ($^\circ$) & & & & &  & &(sec)&\\
        \hline\hline
        2459266.4 & 0.0112 & 96.5 & -0.011 & 0.0014 & -0.0025 & 0.0015 & 6.68 & 0.02 & 7 & 1 \\ \hline
        2459267.4 & 0.0121 & 95.0 & -0.0121 & 0.0017 & -0.0021 & 0.0022 & 6.7 & 0.02 & 10 & 1 \\ \hline
        2459268.4 & 0.0122 & 87.8 & -0.0123 & 0.0019 & 0.001 & 0.0024 & 6.77 & 0.02 & 7 & 1 \\ \hline
        2459269.4 & 0.0126 & 91.5 & -0.0127 & 0.0017 & -0.0007 & 0.0021 & 6.83 & 0.02 & 10 & 1 \\ \hline
        2459270.4 & 0.0106 & 87.8 & -0.0106 & 0.0014 & 0.0008 & 0.0016 & 6.83 & 0.02 & 10 & 1 \\ \hline
        2459271.4 & 0.0097 & 87.1 & -0.0096 & 0.0014 & 0.0010 & 0.0015 & 6.87 & 0.02 & 7 & 1 \\ \hline
        2459273.3 & 0.0133 & 94.0 & -0.0133 & 0.0013 & -0.0018 & 0.0015 & 6.91 & 0.02 & 7& 1 \\ \hline
        2459274.3 & 0.0147 & 86.4 & -0.0149 & 0.0026 & 0.0019 & 0.0027 & 6.87 & 0.02 & 7 & 1 \\ \hline
        2459275.3 & 0.0108 & 86.3 & -0.0109 & 0.0015 & 0.0014 & 0.001 & 6.92 & 0.02 & 10 & 1 \\ \hline
        2459276.3 & 0.0103 & 77.1 & -0.0094 & 0.0015 & 0.0046 & 0.0015 & 6.9 & 0.02 & 10 & 1 \\ \hline
        2459276.4 & 0.0100 & 73.4 & -0.0084 & 0.0008 & 0.0055 & 0.0007 & 6.93 & 0.02 & 10 & 2 \\ \hline
        2459279.3 & 0.0114 & 64.8 & -0.0073 & 0.0009 & 0.0088 & 0.0011 & 6.97 & 0.02 & 10 & 2 \\ \hline
        2459280.3 & 0.0101 & 64.3 & -0.0068 & 0.0041 & 0.0086 & 0.0045 & 6.74 & 0.02 & 10 & 1 \\ \hline
        2459282.3 & 0.0112 & 80.5 & -0.0106 & 0.0012 & 0.0037 & 0.0015 & 7.02 & 0.02 & 10 & 2 \\ \hline
        2459287.4 & 0.0110 & 79.8 & -0.0104 & 0.0012 & 0.0039 & 0.0015 & 7.03 & 0.02 & 10 & 2 \\ \hline
        2459290.4 & 0.0114 & 80.6 & -0.0108 & 0.0009 & 0.0037 & 0.0009 & 7.04 & 0.02 & 12 & 2 \\ \hline
        2459292.4 & 0.0109 & 81.2 & -0.0104 & 0.0005 & 0.0033 & 0.0007 & 7.08 & 0.02 & 10 & 2 \\ \hline
        2459295.4 & 0.0116 & 85.7 & -0.0115 & 0.0005 & 0.0017 & 0.0008 & 7.08 & 0.02 & 10 & 2 \\ \hline
        2459297.3 & 0.0125 & 92.3 & -0.0124 & 0.0004 & -0.001 & 0.0016 & 7.1 & 0.02 & 10 & 1 \\ \hline
        2459298.3 & 0.0107 & 89.7 & -0.0107 & 0.0005 & 0.0001 & 0.0006 & 7.11 & 0.02 & 10 & 1 \\ \hline
        2459303.3 & 0.0125 & 95.1 & -0.0123 & 0.0006 & -0.0022 & 0.0007 & 7.35 & 0.02 & 10 & 1 \\ \hline
        2459304.3 & 0.0134 & 96.6 & -0.013 & 0.0006 & -0.003 & 0.0006 & 7.26 & 0.02 & 10 & 1 \\ \hline
        2459305.4 & 0.0133 & 92.3 & -0.0133 & 0.001 & -0.001 & 0.0007 & 7.17 & 0.02 & 5 & 1 \\ \hline
        2459307.3 & 0.0149 & 98.3 & -0.0143 & 0.0003 & -0.0043 & 0.0007 & 7.23 & 0.02 & 5 & 1 \\ \hline
        2459308.4 & 0.0161 & 94.4 & -0.0159 & 0.0006 & -0.0025 & 0.0007 & 7.2 & 0.02 & 10 & 2 \\ \hline
        2459309.3 & 0.0168 & 101.1 & -0.0156 & 0.0008 & -0.0064 & 0.0007 & 7.2 & 0.02 & 5 & 1 \\ \hline
        2459313.3 & 0.0203 & 104.0 & -0.018 & 0.0006 & -0.0096 & 0.0007 & 7.35 & 0.02 & 5 & 1 \\ \hline
        2459314.4 & 0.0232 & 103.8 & -0.0206 & 0.0006 & -0.0108 & 0.0009 & - & - & 5 & 1 \\ \hline
        2459316.3 & 0.0249 & 105.8 & -0.0212 & 0.0003 & -0.0131 & 0.0007 & 7.25 & 0.02 & 5 & 1 \\ \hline
        2459317.4 & 0.0234 & 99.6 & -0.0221 & 0.0005 & -0.0077 & 0.0007 & 7.25 & 0.02 & 10 & 2 \\ \hline
        2459319.4 & 0.0248 & 99.9 & -0.0233 & 0.0005 & -0.0084 & 0.0007 & 7.28 & 0.02 & 11 & 2 \\ \hline
        2459320.4 & 0.0241 & 100.7 & -0.0225 & 0.0006 & -0.0087 & 0.0007 & 7.29 & 0.02 & 11 & 2 \\ \hline
        2459321.3 & 0.0242 & 106.5 & -0.0203 & 0.0007 & -0.0132 & 0.0007 & 7.36 & 0.02 & 5 & 1 \\ \hline
        2459325.3 & 0.0258 & 108.7 & -0.0205 & 0.0011 & -0.0157 & 0.0013 & 7.44 & 0.02 & 5 & 1 \\ \hline
        2459326.4 & 0.0272 & 103.8 & -0.0241 & 0.0007 & -0.0126 & 0.0009 & 7.4 & 0.02 & 10 & 2 \\ \hline
        2459327.3 & 0.0241 & 108.3 & -0.0194 & 0.0006 & -0.0144 & 0.0007 & 7.43 & 0.02 & 5 & 1 \\ \hline
        2459328.3 & 0.0249 & 109.5 & -0.0193 & 0.0007 & -0.0157 & 0.0007 & 7.48 & 0.02 & 5 & 1 \\ \hline
        2459329.3 & 0.0255 & 110.1 & -0.0195 & 0.0015 & -0.0165 & 0.0009 & 7.6 & 0.02 & 5 & 1 \\ \hline
        2459330.3 & 0.0255 & 110.2 & -0.0194 & 0.0008 & -0.0165 & 0.0008 & 7.55 & 0.02 & 5 & 1 \\ \hline
        2459331.3 & 0.0263 & 108.9 & -0.0208 & 0.0007 & -0.0161 & 0.0007 & 7.54 & 0.02 & 5 & 1 \\ \hline
        2459332.3 & 0.0253 & 109.6 & -0.0196 & 0.0008 & -0.016 & 0.0007 & 7.55 & 0.02 & 5 & 1 \\ \hline
        2459338.3 & 0.0300 & 110.4 & -0.0227 & 0.001 & -0.0197 & 0.0007 & 7.78 & 0.02 & 5 & 1 \\ \hline
        2459340.4 & 0.0235 & 110.6 & -0.0177 & 0.0006 & -0.0155 & 0.0009 & - & - & 5 & 1 \\ \hline
        2459340.4 & 0.0252 & 105.4 & -0.0217 & 0.0009 & -0.0129 & 0.0009 & 7.74 & 0.02 & 15 & 2 \\ \hline
        2459340.5 & 0.0258 & 106.3 & -0.0221 & 0.0011 & -0.0134 & 0.0009 & 7.74 & 0.02 & 15 & 2 \\ \hline
        2459342.5 & 0.0257 & 106.3 & -0.0217 & 0.0009 & -0.0138 & 0.001 & 7.78 & 0.02 & 15 & 2 \\ \hline
        2459342.3 & 0.0257 & 109.5 & -0.0199 & 0.0005 & -0.0162 & 0.0007 & 7.75 & 0.02 & 5 & 1 \\ \hline
        2459343.4 & 0.0212 & 112.9 & -0.0148 & 0.0006 & -0.0152 & 0.0008 & 7.79 & 0.02 & 5 & 1 \\ \hline
        2459353.4 & 0.0190 & 104.6 & -0.0167 & 0.0028 & -0.0094 & 0.0008 & 7.92 & 0.02 & 15 & 2 \\ \hline
        2459354.3 & 0.0212 & 114.0 & -0.0143 & 0.0028 & -0.0159 & 0.0011 & 7.99 & 0.02 & 5 & 1 \\ \hline
        2459364.4 & 0.0117 & 109.7 & -0.0091 & 0.0005 & -0.0074 & 0.0011 & 7.72 & 0.02 & 5 & 1 \\ \hline
        2459365.4 & 0.0140 & 117.6 & -0.008 & 0.0006 & -0.0115 & 0.0007 & 7.57 & 0.02 & 5 & 1 \\ \hline
    \end{tabular}
    }
\end{table*}

\begin{table*}[!ht]
\caption{\label{tab:journal_2021-2}
Journal of measurements from 2021, part 2.}
\centering
    {\renewcommand{\arraystretch}{1.1}
  \begin{tabular}{|ccccccccccc|}
    \hline
    JD & $P_{cal}$ & $PA_{cal}$ & $Q_{cal}$ & $\sigma_Q$ & $U_{cal}$ & $\sigma_{U}$ & $mag(V)$ & $\sigma_{mag(V)}$ & $Exp.$ & $Ob.$ \\
    (d) & (debiased)& ($^\circ$) & & & & &  & &(sec)&\\
    \hline\hline
        2459367.4 & 0.0120 & 111.6 & -0.0088 & 0.0006 & -0.0082 & 0.0008 & 7.57 & 0.02 & 5 & 1 \\ \hline
        2459370.4 & 0.0118 & 106.5 & -0.0099 & 0.0008 & -0.0064 & 0.0010 & 7.66 & 0.02 & 5 & 1 \\ \hline
        2459374.4 & 0.0104 & 106.3 & -0.0088 & 0.0011 & -0.0056 & 0.0012 & 7.38 & 0.02 & 5 & 1 \\ \hline
        2459377.4 & 0.0116 & 104.4 & -0.0102 & 0.0006 & -0.0056 & 0.0009 & 7.51 & 0.02 & 5 & 1 \\ \hline
        2459379.4 & 0.0111 & 102.1 & -0.0101 & 0.0006 & -0.0046 & 0.0008 & 7.16 & 0.02 & 5 & 1 \\ \hline
        2459380.4 & 0.0101 & 111.6 & -0.0088 & 0.002 & -0.0053 & 0.0007 & 7.1 & 0.02 & 5 & 1 \\ \hline
        2459381.4 & 0.0120 & 102.0 & -0.011 & 0.0007 & -0.0049 & 0.0009 & 7.15 & 0.02 & 5 & 1 \\ \hline
        2459383.4 & 0.0112 & 99.8 & -0.0105 & 0.0006 & -0.0038 & 0.0007 & 7.18 & 0.02 & 5 & 1 \\ \hline
        2459389.4 & 0.0123 & 100.8 & -0.0115 & 0.0006 & -0.0045 & 0.0008 & 7.09 & 0.02 & 5 & 1 \\ \hline
        2459392.4 & 0.0135 & 98.4 & -0.013 & 0.0006 & -0.0039 & 0.0008 & 6.88 & 0.02 & 5 & 1 \\ \hline
        2459401.4 & 0.0120 & 98.3 & -0.0115 & 0.001 & -0.0035 & 0.0009 & 7.07 & 0.02 & 5 & 1 \\ \hline
        2459405.4 & 0.0128 & 99.2 & -0.0122 & 0.0007 & -0.004 & 0.0011 & 7.35 & 0.02 & 5 & 1 \\ \hline
        2459407.3 & 0.0140 & 98.7 & -0.0134 & 0.001 & -0.0042 & 0.0007 & 7.31 & 0.02 & 5 & 1 \\ \hline
        2459412.3 & 0.0133 & 97.7 & -0.0129 & 0.0013 & -0.0035 & 0.0012 & 7.31 & 0.02 & 5 & 1 \\ \hline
        2459412.5 & 0.0133 & 93.6 & -0.0132 & 0.0007 & -0.0017 & 0.0008 & 7.35 & 0.02 & 10 & 2 \\ \hline
        2459413.5 & 0.0132 & 93.3 & -0.0131 & 0.0008 & -0.0015 & 0.0009 & 7.36 & 0.02 & 10 & 2 \\ \hline
        2459414.4 & 0.0145 & 99.3 & -0.0138 & 0.0007 & -0.0046 & 0.0014 & 7.34 & 0.02 & 5 &  1 \\ \hline
        2459415.3 & 0.0112 & 99.1 & -0.0107 & 0.0008 & -0.0035 & 0.0009 & 7.44 & 0.02 & 5 & 1 \\ \hline
        2459416.3 & 0.0112 & 96.6 & -0.0109 & 0.0009 & -0.0026 & 0.0009 & 7.3 & 0.02 & 5 & 1 \\ \hline
        2459416.5 & 0.0128 & 92.2 & -0.0128 & 0.0006 & -0.001 & 0.0008 & 7.31 & 0.02 & 10 &2 \\ \hline
        2459417.3 & 0.0108 & 95.6 & -0.0107 & 0.001 & -0.0021 & 0.0008 & 7.25 & 0.02 & 5 & 1 \\ \hline
        2459417.6 & 0.0107 & 91.3 & -0.0108 & 0.0013 & -0.0003 & 0.0013 & 7.33 & 0.02 & 10 & 2 \\ \hline
        2459418.3 & 0.0111 & 95.6 & -0.011 & 0.0012 & -0.0022 & 0.0013 & 7.3 & 0.02 & 5 & 1 \\ \hline
        2459419.3 & 0.0086 & 93.2 & -0.0086 & 0.0013 & -0.001 & 0.0012 & 7.26 & 0.02 & 5 & 1 \\ \hline
        2459425.3 & 0.0078 & 89.2 & -0.0078 & 0.0006 & 0.0002 & 0.0009 & 7.16 & 0.02 & 5 & 1 \\ \hline
        2459429.5 & 0.0063 & 85.9 & -0.0063 & 0.0005 & 0.0009 & 0.0007 & 6.98 & 0.02 & 12 & 2 \\ \hline
        2459436.3 & 0.0043 & 66.9 & -0.003 & 0.001 & 0.0032 & 0.001 & 6.76 & 0.02 & 5 & 1 \\ \hline
        2459437.3 & 0.0037 & 63.0 & -0.0023 & 0.0013 & 0.0032 & 0.0012 & 6.69 & 0.02 & 5 & 1 \\ \hline
        2459439.3 & 0.0033 & 69.4 & -0.0026 & 0.001 & 0.0023 & 0.0011 & 6.71 & 0.02 & 2 & 1 \\ \hline
        2459440.3 & 0.0039 & 67.3 & -0.0029 & 0.0014 & 0.003 & 0.001 & 6.77 & 0.02 & 3 & 1 \\ \hline
        2459441.3 & 0.0046 & 67.5 & -0.0033 & 0.0007 & 0.0033 & 0.0011 & 6.87 & 0.02 & 3 &  1 \\ \hline
        2459453.4 & 0.0052 & 81.6 & -0.005 & 0.0005 & 0.0015 & 0.0007 & 6.77 & 0.02 & 12 & 2 \\ \hline
        2459454.4 & 0.0048 & 79.9 & -0.0045 & 0.0005 & 0.0017 & 0.0007 & 6.8 & 0.02 & 12 & 2 \\ \hline
        2459459.4 & 0.0046 & 86.0 & -0.0046 & 0.0005 & 0.0006 & 0.0006 & 6.86 & 0.02 & 15 & 2 \\ \hline
        2459465.4 & 0.0075 & 84.1 & -0.0074 & 0.0005 & 0.0015 & 0.0006 & 6.93 & 0.02 & 15 & 2 \\ \hline
        2459477.4 & 0.0080 & 100.6 & -0.0075 & 0.0004 & -0.0029 & 0.0006 & 7.1 & 0.02 & 15 & 2 \\ \hline
        2459487.4 & 0.0097 & 99.8 & -0.0092 & 0.001 & -0.0032 & 0.0013 & 7.23 & 0.02 & 15 & 2 \\ \hline
    \end{tabular}
    }
\end{table*}


\begin{table*}[!ht]
\caption{\label{tab:journal_2022-1}
Journal of measurements from 2022, part 1.}
\centering
    {\renewcommand{\arraystretch}{1.1}
   \begin{tabular}{|ccccccccccc|}
    \hline
    JD & $P_{cal}$ & $PA_{cal}$ & $Q_{cal}$ & $\sigma_Q$ & $U_{cal}$ & $\sigma_{U}$ & $mag(V)$ & $\sigma_{mag(V)}$ & $Exp.$ & $Ob.$ \\
    (d) & (debiased)& ($^\circ$) & & & & &  & &(sec)&\\
    \hline\hline
        2459609.4 & 0.0140 & 91.7 & -0.014 & 0.0005 & -0.0008 & 0.0006 & 7.36 & 0.02 & 15 & 2 \\ \hline
        2459617.4 & 0.0120 & 91.4 & -0.012 & 0.0004 & -0.0006 & 0.0006 & 7.01 & 0.02 & 15 & 2 \\ \hline
        2459619.4 & 0.0103 & 93.0 & -0.0103 & 0.0003 & -0.0011 & 0.0006 & 6.89 & 0.02 & 5 & 1 \\ \hline
        2459620.4 & 0.0092 & 89.6 & -0.0092 & 0.0004 & 0.0001 & 0.0007 & 6.82 & 0.02 & 5 & 1 \\ \hline
        2459622.4 & 0.0099 & 90.7 & -0.01 & 0.0008 & -0.0002 & 0.001 & 6.79 & 0.02 & 5 & 1 \\ \hline
        2459623.4 & 0.0089 & 78.7 & -0.0083 & 0.0004 & 0.0034 & 0.0007 & 6.78 & 0.02 & 5 & 1 \\ \hline
        2459624.3 & 0.0088 & 90.0 & -0.0088 & 0.0003 & 0.0000 & 0.0006 & 6.67 & 0.02 & 5 & 1 \\ \hline
        2459625.4 & 0.0090 & 91.2 & -0.009 & 0.0004 & -0.0004 & 0.0006 & 6.68 & 0.02 & 15 & 2 \\ \hline
        2459626.4 & 0.0102 & 90.6 & -0.0102 & 0.0005 & -0.0002 & 0.0007 & 6.61 & 0.02 & 5 & 1 \\ \hline
        2459630.4 & 0.0090 & 94.1 & -0.0089 & 0.0003 & -0.0013 & 0.0006 & 6.58 & 0.02 & 5 & 1 \\ \hline
        2459634.4 & 0.0080 & 90.2 & -0.008 & 0.0004 & 0.0000 & 0.0007 & 6.57 & 0.02 & 5 & 1 \\ \hline
        2459637.4 & 0.0078 & 93.3 & -0.0077 & 0.0004 & -0.0009 & 0.0007 & 6.52 & 0.02 & 5 & 1 \\ \hline
        2459637.4 & 0.0092 & 95.3 & -0.009 & 0.0004 & -0.0017 & 0.0006 & 6.54 & 0.02 & 15 & 2 \\ \hline
        2459639.4 & 0.0076 & 90.9 & -0.0077 & 0.0005 & -0.0002 & 0.0007 & 6.63 & 0.02 & 5 & 1 \\ \hline
        2459640.3 & 0.0080 & 93.1 & -0.008 & 0.0003 & -0.0009 & 0.0006 & 6.62 & 0.02 & 5 & 1 \\ \hline
        2459640.4 & 0.0074 & 90.7 & -0.0074 & 0.0004 & -0.0002 & 0.0006 & 6.56 & 0.02 & 14 & 2 \\ \hline
        2459642.4 & 0.0064 & 88.8 & -0.0065 & 0.0003 & 0.0003 & 0.0006 & 6.57 & 0.02 & 5 & 1 \\ \hline
        2459644.4 & 0.0071 & 93.0 & -0.0071 & 0.0004 & -0.0007 & 0.0006 & 6.61 & 0.02 & 15 & 2 \\ \hline
        2459646.3 & 0.0076 & 99.9 & -0.0072 & 0.0004 & -0.0026 & 0.0006 & 6.61 & 0.02 & 13 & 2 \\ \hline
        2459653.5 & 0.0093 & 94.0 & -0.0092 & 0.0004 & -0.0013 & 0.0006 & 6.68 & 0.02 & 10 &  2 \\ \hline
        2459657.4 & 0.0067 & 80.4 & -0.0063 & 0.0004 & 0.0022 & 0.0006 & 6.76 & 0.02 & 10 & 2 \\ \hline
        2459658.4 & 0.0083 & 92.5 & -0.0083 & 0.0004 & -0.0007 & 0.0006 & 6.77 & 0.02 & 10 & 2 \\ \hline
        2459659.4 & 0.0085 & 93.6 & -0.0084 & 0.0005 & -0.0011 & 0.0006 & 6.79 & 0.02 & 15 & 2 \\ \hline
        2459661.4 & 0.0100 & 96.0 & -0.0098 & 0.0005 & -0.0021 & 0.0007 & 6.83 & 0.02 & 15 & 2 \\ \hline
        2459664.4 & 0.0101 & 96.9 & -0.0098 & 0.0004 & -0.0024 & 0.0006 & 6.92 & 0.02 & 13 & 2 \\ \hline
        2459666.4 & 0.0090 & 89.1 & -0.009 & 0.0005 & 0.0003 & 0.0006 & 6.95 & 0.02 & 11 & 2 \\ \hline
        2459669.4 & 0.0128 & 91.2 & -0.0128 & 0.0005 & -0.0005 & 0.0007 & 7.01 & 0.02 & 11 & 2 \\ \hline
        2459679.3 & 0.0171 & 98.2 & -0.0164 & 0.0021 & -0.0048 & 0.0021 & 7.24 & 0.02 & 5 & 1 \\ \hline
        2459681.3 & 0.0172 & 101.3 & -0.0159 & 0.0003 & -0.0066 & 0.0007 & 7.29 & 0.02 & 5 & 1 \\ \hline
        2459683.3 & 0.0171 & 98.4 & -0.0164 & 0.0004 & -0.0049 & 0.0007 & 7.23 & 0.02 & 5 & 1 \\ \hline
        2459684.3 & 0.0158 & 94.5 & -0.0157 & 0.0005 & -0.0025 & 0.0008 & 7.28 & 0.02 & 5 & 1 \\ \hline
        2459686.3 & 0.0180 & 103.5 & -0.0161 & 0.0006 & -0.0082 & 0.0008 & 7.24 & 0.02 & 5 & 1 \\ \hline
        2459687.3 & 0.0186 & 105.1 & -0.0161 & 0.0005 & -0.0094 & 0.0007 & 7.21 & 0.02 & 5 & 1 \\ \hline
        2459688.4 & 0.0182 & 105.7 & -0.0156 & 0.0005 & -0.0095 & 0.0007 & 7.24 & 0.02 & 5 & 1 \\ \hline
        2459691.3 & 0.0184 & 110.7 & -0.0138 & 0.0016 & -0.0122 & 0.0017 & 7.27 & 0.02 & 5 & 1 \\ \hline
        2459692.4 & 0.0190 & 105.6 & -0.0163 & 0.0004 & -0.0098 & 0.0006 & 7.22 & 0.02 & 12 & 2 \\ \hline
        2459697.4 & 0.0166 & 107.9 & -0.0135 & 0.0004 & -0.0097 & 0.0007 & 7.2 & 0.02 & 12 & 2 \\ \hline
        2459698.4 & 0.0157 & 115.2 & -0.0101 & 0.0021 & -0.0122 & 0.0022 & 7.2 & 0.02 & 5 & 1 \\ \hline
        2459709.4 & 0.0107 & 114.1 & -0.0071 & 0.0005 & -0.008 & 0.0008 & 7.16 & 0.02 & 5 & 1 \\ \hline
        2459715.3 & 0.0104 & 109.5 & -0.0081 & 0.0006 & -0.0065 & 0.0008 & 7.06 & 0.02 & 5 & 1 \\ \hline
        2459717.3 & 0.0095 & 108.3 & -0.0076 & 0.0004 & -0.0057 & 0.0007 & 7.22 & 0.02 & 5 & 1 \\ \hline
        2459718.3 & 0.0092 & 107.4 & -0.0076 & 0.0005 & -0.0053 & 0.0007 & 7.29 & 0.02 & 5 & 1 \\ \hline
        2459721.3 & 0.0082 & 103.2 & -0.0073 & 0.0006 & -0.0036 & 0.0008 & 7.37 & 0.02 & 5 & 1 \\ \hline
        2459725.3 & 0.0117 & 106.8 & -0.0098 & 0.0007 & -0.0065 & 0.0009 & 7.41 & 0.02 & 5 & 1 \\ \hline
    \end{tabular}
    }
\end{table*}

\begin{table*}[!ht]
\caption{\label{tab:journal_2022-2}
Journal of measurements from 2022, part 2.}
\centering
     {\renewcommand{\arraystretch}{1.1}
    \begin{tabular}{|ccccccccccc|}
    \hline
    JD & $P_{cal}$ & $PA_{cal}$ & $Q_{cal}$ & $\sigma_Q$ & $U_{cal}$ & $\sigma_{U}$ & $mag(V)$ & $\sigma_{mag(V)}$ & $Exp.$ & $Ob.$ \\
       (d) & (debiased)& ($^\circ$) & & & & &  & &(sec)&\\
        \hline\hline
        2459726.4 & 0.0128 & 109.5 & -0.01 & 0.0007 & -0.0081 & 0.0009 & 7.45 & 0.02 & 5 & 1 \\ \hline
        2459730.4 & 0.0124 & 104.2 & -0.011 & 0.0005 & -0.0059 & 0.0008 & 7.54 & 0.02 & 5 & 1 \\ \hline
        2459733.4 & 0.0137 & 102.8 & -0.0124 & 0.0007 & -0.006 & 0.0009 & 7.59 & 0.02 & 5 & 1 \\ \hline
        2459741.4 & 0.0175 & 90.0 & -0.0175 & 0.0005 & 0.0000 & 0.0007 & 7.71 & 0.02 & 5 & 1 \\ \hline
        2459742.4 & 0.0176 & 101.2 & -0.0163 & 0.0007 & -0.0067 & 0.0009 & 7.71 & 0.02 & 5 & 1 \\ \hline
        2459743.4 & 0.0183 & 99.9 & -0.0172 & 0.0007 & -0.0062 & 0.0009 & 7.79 & 0.02 & 5 & 1 \\ \hline
        2459745.4 & 0.0188 & 99.1 & -0.0179 & 0.001 & -0.0059 & 0.0011 & 7.74 & 0.02 & 5 & 1 \\ \hline
        2459750.4 & 0.0199 & 93.3 & -0.0198 & 0.0009 & -0.0023 & 0.001 & 7.76 & 0.02 & 5 & 1 \\ \hline
        2459760.3 & 0.0198 & 97.8 & -0.0191 & 0.0012 & -0.0053 & 0.0013 & 7.63 & 0.02 & 5 & 1 \\ \hline
        2459762.4 & 0.0153 & 90.5 & -0.0153 & 0.0006 & -0.0003 & 0.0008 & 7.58 & 0.02 & 5 & 1 \\ \hline
        2459763.4 & 0.0228 & 94.4 & -0.0226 & 0.0007 & -0.0035 & 0.0009 & 7.58 & 0.02 & 5 & 1 \\ \hline
        2459764.4 & 0.0150 & 84.2 & -0.0147 & 0.0009 & 0.003 & 0.001 & 7.47 & 0.02 & 5 & 1 \\ \hline
        2459766.4 & 0.0193 & 92.5 & -0.0193 & 0.0006 & -0.0017 & 0.0008 & 7.53 & 0.02 & 5 & 1 \\ \hline
        2459769.4 & 0.0188 & 94.8 & -0.0186 & 0.0005 & -0.0031 & 0.0007 & 7.52 & 0.02 & 5 & 1 \\ \hline
        2459771.4 & 0.0172 & 90.0 & -0.0172 & 0.0002 & 0.0000 & 0.0006 & 7.54 & 0.02 & 5 & 1 \\ \hline
        2459773.3 & 0.0189 & 93.7 & -0.0188 & 0.0017 & -0.0025 & 0.0018 & 7.52 & 0.02 & 5 & 1 \\ \hline
        2459777.3 & 0.0177 & 91.8 & -0.0177 & 0.0009 & -0.0011 & 0.0011 & 7.46 & 0.02 & 5 & 1 \\ \hline
        2459778.3 & 0.0182 & 93.4 & -0.0181 & 0.0006 & -0.0022 & 0.0008 & 7.45 & 0.02 & 5 & 1 \\ \hline
        2459780.3 & 0.0170 & 94.8 & -0.0168 & 0.0005 & -0.0028 & 0.0007 & 7.43 & 0.02 & 5 & 1 \\ \hline
        2459782.3 & 0.0181 & 93.3 & -0.018 & 0.001 & -0.0021 & 0.0012 & 7.38 & 0.02 & 5 & 1 \\ \hline
        2459785.3 & 0.0182 & 95.7 & -0.0179 & 0.0007 & -0.0036 & 0.0009 & 7.32 & 0.02 & 5 & 1 \\ \hline
        2459787.3 & 0.0155 & 93.5 & -0.0154 & 0.0013 & -0.0019 & 0.0014 & 7.28 & 0.02 & 5 & 1 \\ \hline
        2459794.3 & 0.0157 & 96.7 & -0.0153 & 0.0005 & -0.0036 & 0.0007 & 7.19 & 0.02 & 5 & 1 \\ \hline
        2459795.3 & 0.0182 & 100.6 & -0.017 & 0.0005 & -0.0066 & 0.0008 & 7.2 & 0.02 & 5 & 1 \\ \hline
        2459796.3 & 0.0178 & 98.5 & -0.0171 & 0.0006 & -0.0052 & 0.0008 & 7.16 & 0.02 & 5 & 1 \\ \hline
        2459800.3 & 0.0175 & 99.2 & -0.0166 & 0.0008 & -0.0055 & 0.001 & 7.2 & 0.02 & 5 & 1 \\ \hline
        2459801.3 & 0.0148 & 97.0 & -0.0144 & 0.0007 & -0.0036 & 0.0009 & 7.21 & 0.02 & 5 & 1 \\ \hline
        2459802.3 & 0.0159 & 99.0 & -0.0151 & 0.0007 & -0.0049 & 0.0009 & 7.28 & 0.02 & 5 & 1 \\ \hline
        2459802.5 & 0.0186 & 94.2 & -0.0184 & 0.0004 & -0.0027 & 0.0006 & 7.22 & 0.02 & 12 & 2 \\ \hline
        2459803.4 & 0.0161 & 94.5 & -0.0159 & 0.0006 & -0.0024 & 0.0007 & 7.22 & 0.02 & 11 & 2 \\ \hline
        2459808.3 & 0.0164 & 97.6 & -0.0158 & 0.0008 & -0.0043 & 0.0009 & 7.25 & 0.02 & 5 & 1 \\ \hline
        2459810.5 & 0.0175 & 95.6 & -0.0172 & 0.0007 & -0.0034 & 0.0007 & 7.25 & 0.02 & 14 & 2 \\ \hline
        2459815.4 & 0.0160 & 95.9 & -0.0157 & 0.0009 & -0.0033 & 0.001 & 7.2 & 0.02 & 14 & 2 \\ \hline
        2459824.4 & 0.0164 & 100.6 & -0.0153 & 0.0005 & -0.0059 & 0.0006 & 7.17 & 0.02 & 13 & 2 \\ \hline
        2459833.4 & 0.0145 & 97.1 & -0.0141 & 0.0006 & -0.0036 & 0.0008 & 7.21 & 0.02 & 15 & 2 \\ \hline
        2459837.4 & 0.0112 & 95.4 & -0.011 & 0.0005 & -0.0021 & 0.0007 & 7.17 & 0.02 & 15 & 2 \\ \hline
        2459845.3 & 0.0107 & 100 & -0.0101 & 0.0005 & -0.0037 & 0.0007 & 7.12 & 0.02 & 15 & 2 \\ \hline
        2459847.3 & 0.0116 & 97.8 & -0.0112 & 0.0004 & -0.0031 & 0.0007 & 7.12 & 0.02 & 15 & 2 \\ \hline
        2459849.3 & 0.0132 & 95.2 & -0.013 & 0.0005 & -0.0024 & 0.0006 & 7.1 & 0.02 & 15 & 2 \\ \hline
        2459850.3 & 0.0128 & 103.7 & -0.0114 & 0.0005 & -0.0059 & 0.0006 & 7.08 & 0.02 & 15 & 2 \\ \hline
        2459865.4 & 0.0173 & 111.5 & -0.0126 & 0.0004 & -0.0118 & 0.0007 & 6.98 & 0.02 & 15 & 2 \\ \hline
        2459867.3 & 0.0169 & 111.9 & -0.0122 & 0.0005 & -0.0117 & 0.0007 & 7.00 & 0.02 & 15 & 2 \\ \hline
        2459887.7 & 0.0184 & 124.0 & -0.0069 & 0.0006 & -0.0171 & 0.0008 & 7.33 & 0.02 & 15 & 2 \\ \hline
    \end{tabular}
    }
\end{table*}
\end{appendix}

\end{document}